\newif\ifabstract
\abstracttrue
 \abstractfalse 
\newif\iffull
\ifabstract \fullfalse \else \fulltrue \fi

\documentclass[11pt]{article}
\usepackage{amsfonts}
\usepackage{amssymb}
\usepackage{amstext}
\usepackage{amsmath}
\usepackage{xspace}
\usepackage{theorem}
\usepackage{graphicx}
\usepackage{url}
\usepackage{graphics}
\usepackage{colordvi}
\usepackage{colordvi}
\usepackage{subfigure}

\usepackage{natbib}
\usepackage{wasysym}

\usepackage{moreverb} 
\usepackage{textcomp} \usepackage{epsfig}
\usepackage{color} \usepackage{upgreek}
\usepackage{bbm}

\usepackage{subeqnarray}

\usepackage{xcolor}

\advance \oddsidemargin by -0.8in
\newcommand{\myparskip}{3pt}
\parskip \myparskip

\newcommand\Rey{\mbox{\textit{Re}}}  

\newsavebox{\astrutbox}
\sbox{\astrutbox}{\rule[-5pt]{0pt}{20pt}}

\begin{document}

\title{Optimal control of a separated boundary-layer flow over a bump}
\author{Pierre-Yves Passaggia\thanks{Department of Marine Sciences, University of North Carolina, Chapel Hill, NC 27599, USA. Email: passaggia@unc.edu {\tt passaggia@unc.edu}.} \thanks{Aix-Marseille Univ, CNRS, Centrale Marseille, IRPHE, Marseille, France} 
\and Uwe Ehrenstein \thanks{Aix-Marseille Univ, CNRS, Centrale Marseille, IRPHE, Marseille, France Email: {\tt ehrenstein@irphe.univ-mrs.fr}.}}
\maketitle

\thispagestyle{empty}

\begin{abstract}
The optimal control of a globally unstable two-dimensional separated boundary layer over a bump is
considered using augmented Lagrangian optimization procedures. The present strategy allows \Blue{of} 
controlling the flow from a fully developed nonlinear state back to the steady state using a single actuator. 
The method \Blue{makes use of} a decomposition between the slow dynamics associated with the baseflow modification, and the fast dynamics characterized by \Blue{a large scale oscillation of the recirculation region, known as flapping.} Starting from a steady state forced by a suction actuator located near the separation point, the baseflow modification is shown to be controlled by a vanishing suction strategy. 
For weakly unstable flow regimes, this control law can be further optimized by means of direct-adjoint iterations of the nonlinear Navier-Stokes equations. 
In the absence of external noise, this novel 
approach proves to be capable of controlling 
the transient dynamics and the baseflow modification simultaneously.
\end{abstract}


\section{Introduction}

\Blue{Boundary-layer flows subject to an adverse pressure gradient can separate from the boundary and exhibit recirculation regions characterized by closed streamlines. The local properties of these separated flows can trigger different types of instabilities.} 
\Blue{Closed streamlines may induce three-dimensional centrifugal-type
instabilities as well as elliptic instability mechanisms, which have 
been shown for instance for backward-facing step flow \citep*{Barkley:02,Beaudoin:04,Lanzerstorfer:12} or for a flow over a bump 
\citep{gallaire:marquillie:ehrenstein}.}

\Blue{Two-dimensional oscillatory instabilities due to the shear along the 
baseflow separating streamline, that is Kelvin-Helmholtz type instabilities, have been observed for instance in open cavity flows \citep{Sipp:07}, whereas for a separating boundary-layer induced by a bump global two-dimensional oscillations have been predicted in \citet{ehrenstein:gallaire:bosse}. Model separation bubbles have been
addressed for instance in \cite{Rodriguez:13}, questioning the existence of a two-dimensional global oscillator behaviour.}  

\Blue{The existence of an unstable global mode gives rise to a so-called resonator dynamics, contrary to an amplifier dynamics, which is likely to be observed  
in open shear flows even in the absence of individual instability modes. Indeed, the non-normality of the linear stability operator for those baseflows \citep*{schmid:henningson, Schmid:07} may provide a powerful instability mechanism, associated with the non-normal coupling between individually stable (or weakly unstable) modes.}

The optimal control of both the resonator and the amplifier dynamics is a matter of active research \citep{kim:bewley,Cattafesta:11,Duriez:17} and new algorithms, capable of efficiently controlling flows that are both resonators and amplifiers in nature, have yet to be proposed.
Recently \citet*{gautier:aider:2,gautier:aider} successfully implemented an experimental feedback control of the amplifier dynamics in a backward facing step flow, but the control of the resonator dynamics associated with the globally unstable configuration \Blue{for such a flow geometry}
remains an open issue.
Feedback control based on reduced-order models of the linear instability dynamics has received a lot of attention recently \citep{kim:bewley}, 
but proved to lack robustness when considering self-sustained nonlinear dynamics 
\citep*{Huang:08,ehrenstein:passaggia:gallaire,barbagallo:sipp:schmid}. 
\Blue{In the attempt of controlling the nonlinear dynamics in a separated boundary-layer flow over a bump, in \citet{passaggia:ehrenstein:2013} the optimization problem 
has been solved using the augmented Lagrangian approach, the control being determined iteratively by coupling the Navier-Stokes equations with the associated 
adjoint system. While in this approach the unstable dynamics could be attenuated, a definite control could however not be achieved.} 
Another control strategy is to manipulate separated baseflows using Lagrangian-based sensitivity analysis. 
For instance, \citet{boujo:gallaire:2:2014} designed a 
steady actuation approach to modify quantities of interest such as the recirculation length
\Blue{or the geometry of the recirculation bubble's separatrix.}

\Blue{The separated flow case used in the present analysis is the two-dimensional boundary-layer flow over a shallow bump, which has previously been considered in the 
numerical investigation by \citet{marquillie:ehrenstein} and in the experimental study by \citet{passaggia:leweke:ehrenstein}. The control of this separated flow 
case is particularly challenging, given that its instability dynamics exhibits strong transient growth as well as multiple unstable time scales, as shown in  \citet{ehrenstein:gallaire:bosse}. A simple open-loop control procedure for this flow case has been designed in \citet{boujo:ehrenstein:gallaire:2013}, showing that steady wall suction at the bump summit, while 
decreasing the recirculation length, also suppresses the unstable nature of the flow
and suction hence appears as a natural starting point for a control strategy.
In the present investigation, we precisely combine a sensitivity analysis and an optimal control strategy, the flow actuation being achieved by suction at the bump's summit, aiming at controlling the flow back to its steady state for a time vanishing suction amplitude.}

\Blue{The paper is organized as follows. In \S2, the flow geometry and the numerical simulation procedure is 
briefly outlined. The control procedure consists of four steps which  
are outlined as follows. 
\begin{itemize}
 \item The sensitivity analysis is described in \S3, the key quantity for the baseflow modification being the recirculation area.
 \item The step response of the flow is then analyzed in \S4 to determine the modal properties of the slow baseflow modification dynamics.  
 \item An {\it ad-hoc} suction control law for this slow baseflow modification is determined explicitly in \S5.1 by considering a reduced order model of the recirculation area modification.
 \item This suction control law is further optimized in \S5.2 using Lagrangian-based optimal control over small time windows and advancing in time using a receding-predictive algorithm \citep{bewley:moin:temam} to suppress the fast transient dynamics.
\end{itemize}
Results are shown for different Reynolds numbers in \S6 and conclusions are given in \S7.}

\section{Geometry and steady state simulation procedure}

\subsection{Problem setup}

The fluid motion is governed by the two-dimensional 
incompressible Navier-Stokes system 
written in dimensionless form 
\begin{equation}
[\partial_t \mathbf{U}; \; 0]=\mathbf{F}(\mathbf{U},P,\Rey), \; 
\mbox{with} \; \mathbf{F}(\mathbf{U},P,\Rey)=[-(\mathbf{U} \cdot \nabla) \mathbf{U} -\nabla
P +\frac{1}{\Rey}\nabla^{2}\mathbf{U}; \; \nabla\cdot\mathbf{U}]
\label{NSU}
\end{equation}
for the velocity field $\mathbf{U}(x,y,t)$ and the pressure $P(x,y,t)$.
The Reynolds number in (\ref{NSU}) is  $\Rey = \delta^{*}U_{\infty}/\nu$,
where the displacement thickness $\delta^{*}$ of the Blasius profile imposed at inflow $x=0$
is used as reference length. The free stream velocity $U_{\infty}$ is the reference velocity and $\nu$ is the dynamic viscosity.
The two-dimensional flow domain is $0\le x\le L$, $\eta(x)\le y < y_{max}$, 
with $\eta(x)$ the lower boundary containing a bump of height $h$. 
The summit of the bump is localized at $x=25$.
\Blue{According to a previous numerical analysis for this flow geometry  \citep{marquillie:ehrenstein}},
a domain length $L=250$ and upper freestream boundary $y_{max} = 100$ 
\Blue{are convenient}
to minimize finite domain effects.
The freestream condition $(U,V) = (1,0)$ has been imposed whereas no-slip conditions are 
used at the wall $y = \eta(x)$. The stress-free boundary condition
\begin{equation}
\frac{1}{Re} (\nabla \mathbf{U}) \cdot 
\mathbf{n} |_{x = L} = P \mathbf{n} |_{x = L} 
\label{stress:free}
\end{equation}
($\mathbf{n}$ being the unit normal vector at the flow domain's boundary) 
has been implemented at the outflow $x = L$ and it proved to be appropriate 
for both the simulations and the subsequent optimization procedures.
\begin{figure}
\begin{center}
\scalebox{0.85}{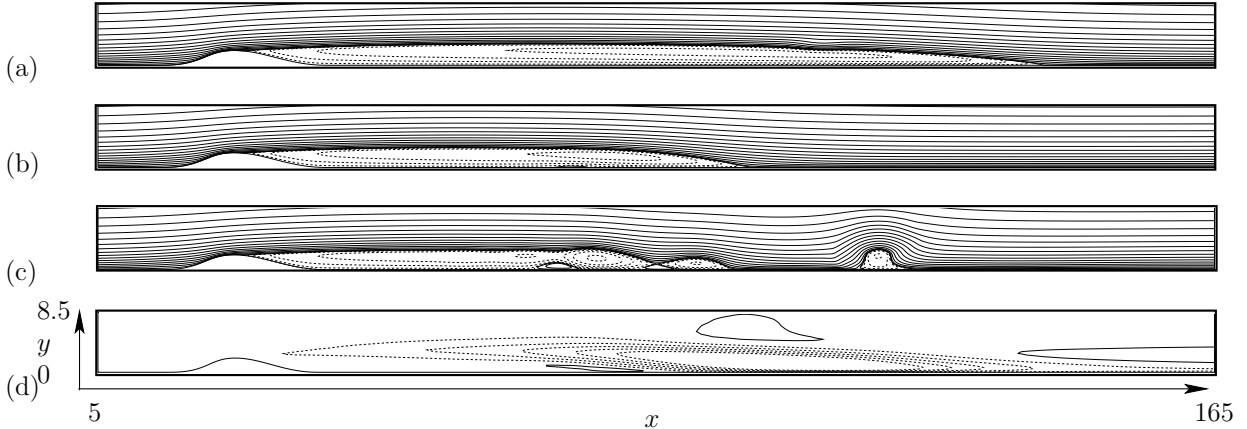}
\caption{Streamlines of the baseflow 
$\mathbf{Q}_0$
(a), of the mean flow $\bar{\mathbf{Q}}$ (b), and of a snapshot of the unstable flow field (c), at $Re=650$ and a bump height $h=2$ (the values of the iso-contours are chosen arbitrarily to represent the flow dynamics). {\Blue{Baseflow modification
$\delta \mathbf{Q} = \mathbf{Q}_0 - \bar{\mathbf{Q}}$ (d), 
shown by isocontours of streamwise velocity $u$ in the range $[-0.35,-0.28,-0.23\dots,0.13]$, the dashed/continuous lines being the negative/positive  values respectively.
}}}
\label{etat_base} 
\end{center}
\end{figure}
The streamwise direction $x$ is discretized using 4$^{th}$ order finite differences, 
whereas in the wall-normal direction $y$ Chebyshev-collocation is considered and
the pressure is solved using the influence matrix technique.
Details about the numerical discretization procedure can be found in 
\citet{marquillie:ehrenstein:numerique} and \citet{passaggia:ehrenstein:2013}.

\subsection{Selective frequency damping method}\label{sec:sfd}
Steady state solutions of the Navier-Stokes system and its adjoint 
are computed using the selective frequency damping technique (SFD) proposed in  
\cite{akervik:brandt:henningson:hoepffner:marxen:schlatter}.
Writing formally the Navier-Stokes equations
$\partial_t \mathbf{Q} = \mathbf{F}(\mathbf{Q},Re)$, the 
method consists in solving the coupled system
\begin{equation}
\left\{
\begin{array}{c}
\partial_t \mathbf{Q} = \mathbf{F}(\mathbf{Q},Re) - \zeta
(\mathbf{Q}-\mathbf{R}) \\ \partial_t \mathbf{R} =
\frac{1}{\Lambda}(\mathbf{Q}-\mathbf{R}).\\
\end{array}
\right.
\label{SFD}
\end{equation}
The coupled system (\ref{SFD}) is integrated in time by computing a sliding time average $\mathbf{R}(t)$ of the flow dynamics $\mathbf{Q}(t)$ and subtracting their difference in the Navier-Stokes equations. 
\Blue{The parameter $\Lambda$ is the time window over which the sliding average $\mathbf{R}$ is computed. The damping factor $\zeta$ is akin to a decay rate and has to be sufficiently large to allow for filtering the instabilities from the solution $\mathbf{Q}$ as the system (\ref{SFD}) is integrated in time.}
Once the above algorithm has converged, that is $\partial_t\mathbf{Q} = \partial_t \mathbf{R} = 0$, the solution $\mathbf{R}=\mathbf{Q}$ is the steady state solution of (\ref{NSU}).
Here we use the procedure of \citet{cunha:passaggia:lazareff} to compute optimal parameters for the SFD.
For instance at $\Rey=650$, the best stabilization strategy was found for $(\zeta,\Lambda)=(0.0292,14.99)$. 
The resulting steady state,
which in the following will be written $\mathbf{Q}_0$,
is shown in figure \ref{etat_base}(a). 
As can be seen, the flow separates at the bump summit and the steady state is characterized by an 
elongated recirculation bubble which reattaches at $x \approx 125$ for the specific supercritical
Reynolds number $Re = 650$ and bump height $h = 2$ chosen.

Note that the critical Reynolds number for the onset of the two-dimensional global unstable dynamics for 
this bump flow geometry has been found to be $Re_c \approx 590$ in the numerical investigation by  
\citet{marquillie:ehrenstein}, which has more recently been confirmed experimentally in \citet{passaggia:leweke:ehrenstein}.\\


\subsection{Steady \& mean state}

The two-dimensional self-sustained instabilities associated with the present flow geometry have already been 
investigated
\citep{marquillie:ehrenstein:numerique,marquillie:ehrenstein,ehrenstein:gallaire:bosse,passaggia:leweke:ehrenstein}.
In particular, it has been shown that in the unstable regime the time-averaged mean flow exhibits a shorter recirculation bubble, the length of which decreasing with increasing Reynolds number.
A numerical simulation of the developed unstable flow at $Re=650$ (figure \ref{etat_base}(c) 
showing a snapshot sample of the unsteady flow field) 
was performed to compute the mean state 
by time averaging
\begin{equation}
\mathbf{\bar{Q}} = \frac{1}{\Delta t}\int_{t}^{t+\Delta t} \mathbf{Q}(\mathbf{x},t) \mbox{d}t,
\end{equation}
with $\Delta t = 4000$ in the above integral. 
The corresponding mean flow is shown in figure \ref{etat_base}(b). 
The difference 
\begin{equation}
\delta \mathbf{Q} = \mathbf{Q}_0 -\mathbf{\bar{Q}}
\end{equation}
between the steady state and the mean flow state plays an important role in the dynamics and $\delta \mathbf{Q}$,
shown in figure \ref{etat_base}(d) for the flow case at $Re = 650$ considered,
will be seen to be a key quantity for 
the control of this flow configuration. 

The effects of a small modification of the steady-state on self-excited flows characterized by a strong dominant frequency was addressed for instance by \citet{Sipp:07} in the case of a separated boundary layer flow over a square cavity and more recently by \citet{Truton:Tuckerman:Barkley} in the case of thermosolutal convection. 
The latter showed that the linearized operator around the mean flow indeed exhibits a purely imaginary eigenvalue. 
This criterion has even been considered by \citet{Mantic-Lugo:Arratia:Gallaire:2014,
Mantic-Lugo:Arratia:Gallaire:2015} to build a self-consistent model of the cylinder flow that predicts both the frequency of the vortex shedding and the amplitude of the nonlinear limit cycle  for Reynolds numbers up to $110$. \citet{flinois:colonius} showed that fully developed non-linear 
flow dynamics of the cylinder flow could be controlled using a direct-adjoint approach where they optimized the control law of blowing and suction actuators located at the walls of the cylinder. 
But the conclusions of these studies cannot be extended to the more general case where a flow presents a broadband spectrum of unstable frequencies. 
The attempts in \citet{passaggia:ehrenstein:2013} to control  
the unstable flow dynamics 
for the present separated boundary layer flow 
led to the conclusion, that 
while
the instabilities of the fully developed regime could be attenuated using a direct-adjoint optimization of the nonlinear dynamics, the flow could however not be controlled back to its steady state. 
\Blue{The complex dynamics involving unstable global modes and transient growth led to increasingly large time optimization windows with increasing Reynolds numbers, 
making the optimization problem hardly tractable in the strongly unstable regime.} 

\Blue{In the following, we revisit the control problem for this challenging 
separated bump flow problem. Our approach will consist in a suction-like actuation to control the steady state modification $\delta\mathbf{Q}$, in combination with a direct-adjoint iterative optimization procedure to minimize transient growth phenomena and global instabilities.}  
\Blue{We first consider the sensitivity analysis of the bump flow for a steady suction-type forcing following the Lagrangian optimization approach of \citet{boujo:gallaire:2:2014}, which was shown to damp and even suppress the self-excited instabilities \citep{boujo:gallaire:1:2014}. We therefore seek to compute first the most sensitive region of the flow to a wall-suction actuator with the aim of maximizing the controllability while minimizing the size of the actuator.}

\section{Sensitivity analysis}\label{sec:sens_an}

\Blue{The baseflow modification $\delta\mathbf{Q}$ shown in figure \ref{etat_base}(d) drastically reduces the length but also the structure of the recirculation region (e.g. the angles of the separatrix near the separation/reattachment points, the amount of backflow). Therefore we
consider here the sensitivity} of the recirculation area $A_{\mbox{rec}}$ which is \Blue{an integral quantity} defined as 
\begin{equation}
A_{\mbox{rec}}(t) = \int_{x_s(t)}^{x_r(t)}\int_{\eta(x)}^{y_{sep}(x,t)} \;\mbox{d}x\; \mbox{d}y.
\label{Arec}
\end{equation} 
where $x_{s,r}$ are the locations of the separation/reattachment point at the wall and $y_{sep}(x , t)$ 
is the height of the separatrix, defined as the streamline 
which connects the stagnation points.
In the remaining of this subsection, the dependence on time for the sensitivity analysis is dropped. This is justified because the actuation $\Phi(t)$ is assumed to be slow 
compared with the time scale of instabilities. This assumption will be supported by the DMD analysis in \S\ref{sec:DMD_step} and
time dependence will be reintroduced later, to minimize the recirculation area modification $\delta A_{\mbox{rec}}$, associated with the baseflow modification $\delta \mathbf{Q}$.
The sensitivity of a steady state to a modification of the recirculation area reduces to a field defined through the first 
order 
modification
$\delta A_{\mbox{rec}}$, 
induced by a small amplitude steady blowing-suction control 
$\delta \Phi$ at the wall $\Gamma_c$
and is given by
\begin{equation}
\delta A_{\mbox{rec}} = (\nabla_{\Phi}A_{\mbox{rec}}| \delta \Phi)
\label{sensi}
\end{equation}
with $( \cdot | \cdot)$
the inner product between $\nabla_{\Phi}A_{\mbox{rec}}$, the gradient of the recirculation area with respect to the suction forcing, \Blue{and the small modification of suction 
amplitude $\delta\Phi$}. 
Note that in the present study there is only one actuator, hence $\Phi$ is simply a scalar quantity and the above inner product is here a simple product between scalars.

\begin{figure}
\hspace{10mm}\scalebox{0.85}{\input{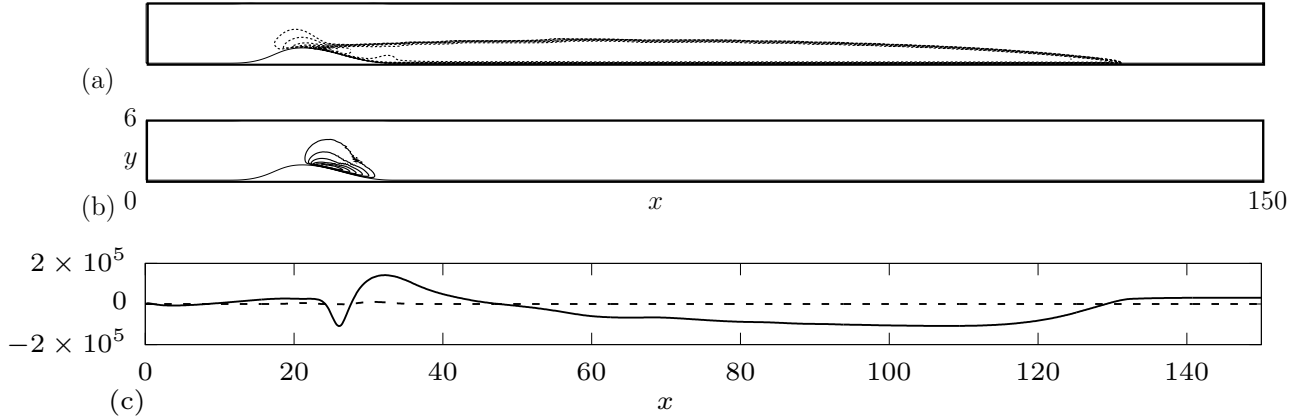}}

\hspace{8mm}\scalebox{1.425}{\input{grad_sens}}
\caption{Contours of the adjoint solutions $U^+$ (a) and $V^+$ (b) for the sensitivity of 
         the recirculation area $A_{rec}$ at $\Rey=650$. 
         The continuous/dotted lines represent positive/negative contours respectively.
         (c) Values of the gradient $-\Rey^{-1}\nabla\mathbf{U}^{+}\cdot\mathbf{n} + P^{+}\mathbf{n}$ evaluated at $\eta_x$ for the first component (- - -) and the second component (-----).}
\label{fig:sensitiv}
\end{figure}

The Lagrange multipliers $\mathbf{Q}^{+}$ are introduced and the Lagrangian
\begin{equation}
\mathcal{L}(\mathbf{Q},\mathbf{Q}^{+}) = A_{\mbox{rec}} \;- <\mathbf{F}(\mathbf{Q},Re),\mathbf{Q}^{+}> \; - \; \int_{\Gamma_{c}} \mathbf{g}(\mathbf{U},\Phi)\cdot \mathbf{B}\Phi^+ \;\mbox{d}\mathbf{s}, 
\label{Lagrangian1}
\end{equation}
is to be rendered stationary. The scalar product $< \cdot , \cdot>$ 
is defined
by the integral $\int_{\Omega} \;\cdot\;\mbox{d}\mathbf{x}$ where $\Omega$ is the flow domain.
The boundary condition for the velocity field at $\Gamma_c$ associated with the control $\Phi$
\begin{equation}
\mathbf{g}(\mathbf{U},\Phi) = \mathbf{U}|_{\Gamma_{c}}-\mathbf{B}\Phi = 0,
\label{boundcontrol}
\end{equation}
is added to the Lagrangian with $\Phi^+$ the Lagrange multiplier associated with the control $\Phi$. The shape-function $\mathbf{B}$ corresponds to the unit vector, orthogonal to $\eta(x)$ along $\Gamma_c$ on the portion of the boundary where the control is applied. The adjoint Navier-Stokes system is 
\Blue{derived from (\ref{Lagrangian1}) 
by taking the Fr\'echet derivative 
defined such that
\begin{equation}
\mathcal{L}_{\mathbf{Q}} (\mathbf{Q} , \mathbf{Q}^+) \cdot \hat{\mathbf{Q}} = \lim_{\epsilon \rightarrow 0} \frac{1}{\epsilon} \left( \mathcal{L} (\mathbf{Q} + \epsilon  \hat{\mathbf{Q}}, \mathbf{Q}^+) - \mathcal{L} (\mathbf{Q},\mathbf{Q}^+) \right)
\label{Frechet}
\end{equation}
with respect to $\mathbf{Q}$ (see \citet{joslin:gunburger:nicolaides:erlebacher:hussaini})
and is given by}
\begin{equation}
\mathbf{F}^+(\mathbf{U},\mathbf{Q}^+,Re)=[ -(\mathbf{U} \cdot \nabla) \mathbf{U}^+ +(\nabla\mathbf{U})^T\mathbf{U}^+ + \nabla
P^+ -\frac{1}{Re}\nabla^{2}\mathbf{U}^+; \; - \nabla\cdot\mathbf{U}^+].
\label{NSUad}
\end{equation}
The boundary terms 
are given by
\begin{equation}
\int_{\partial\Omega} \bigg{[}
\frac{1}{Re}\mathbf{U}^{+}(\nabla\hat{\mathbf{U}}\cdot \mathbf{n})
-\frac{1}{Re}(\nabla\mathbf{U}^{+}\cdot \mathbf{n}) \hat{\mathbf{U}}
-(\mathbf{U}\cdot \mathbf{n})\mathbf{U}^{+}\cdot \hat{\mathbf{U}}
+(P^{+}\cdot\mathbf{n})\hat{\mathbf{U}} - \mathbf{U}^{+}(\hat{P}\cdot
\mathbf{n}) \bigg{]} \; \mbox{d}\mathbf{s},
\label{bord1}
\end{equation}
with $\mathbf{n}$ the unit vector, normal to the boundary $\partial \Omega$. 
The boundary conditions for the velocity field $\mathbf{U}$ are 
\begin{equation} 
\frac{1}{Re} (\nabla \mathbf{U}) \cdot 
\mathbf{n} \Big{|}_{x=L} = P \mathbf{n}\Big{|}_{x=L}, 
\quad \mathbf{U}\Big{|}_{x=0} = (U_{Blas}(y),0),
\quad \mathbf{U}\Big{|}_{y=\eta,H} = 0
\label{bounddirect1}
\end{equation}
and $U_{Blas}$ is the Blasius velocity profile imposed at the inlet.
The same conditions hold for all possible variations in $\hat{\mathbf{U}}, 
\hat{P}$ in the Fr\'echet derivatives and the boundary integral in (\ref{bord1}) vanishes when using the boundary condition for the adjoint system 
\begin{equation}
 \frac{1}{Re}(\nabla\mathbf{U}^{+})\mathbf{n}
+ (\mathbf{U}\cdot \mathbf{n})\mathbf{U}^{+}\Big{|}_{x=0,L}
= P^{+}\mathbf{n}\Big{|}_{x=0,L}, \quad\quad\\
\mathbf{U}^{+}\Big{|}_{y=\eta,H}=0.
\label{boundadjoint1}
\end{equation}
\Blue{The adjoint system (\ref{NSUad}) is subject to a source term related to the $A_{\mbox{rec}}$ term in (\ref{Lagrangian1}). The flow rate through any vertical cross section of the recirculation region is zero and  using this property, \citet{boujo:gallaire:2:2014} showed that}
\begin{equation}
<\nabla_{\mathbf{Q}} A_{\mbox{rec}} , 
\hat{\mathbf{Q}} >
= \int_{x_s}^{x_r} \frac{-1}{U(x,y_{sep}(x))} \left(\int_{y_{w(x)}}^{y_{sep(x)}}
\hat{U}
\mbox{d}y \right)\mbox{d}x,
\end{equation}
\Blue{and the adjoint system to solve is}
\begin{equation}
\mathbf{F}^{+}(\mathbf{U},\mathbf{Q}^{+},Re) = [h(U);0;0],\quad \mbox{with} \quad
h(U)=\int_{x_s}^{x_r} \frac{-1}{U(x,y_{sep}(x))} \left(\int_{y_{w(x)}}^{y_{sep(x)}}\mbox{d}y \right)\mbox{d}x. 
\end{equation}
Similar to the procedure used for the solution of (\ref{NSU}), this adjoint 
system subject to the conditions (\ref{boundadjoint1}) 
is solved using the influence matrix technique outlined in \citet{passaggia:ehrenstein:2013}. The 
steady state adjoint solution $\mathbf{Q}^+$ is obtained through 
time-marching of (\ref{NSUad}), coupled to the selective frequency damping method (cf. \S2).
The adjoint solution for $\mathbf{U}^+$ is shown in figure \ref{fig:sensitiv} at $Re=650$,
where the most sensitive region to blowing or suction is identified in the vicinity of the separation point.

There are remaining terms in the boundary integral (\ref{bord1}) and according to 
the Fr\'echet derivative of the third term in the Lagrangian (\ref{Lagrangian1}), by taking into account (\ref{boundcontrol}),
\Blue{these terms cancel when}
\begin{equation}
\mathbf{B}\Phi^{+} = -\frac{1}{Re}\nabla\mathbf{U}^{+}\cdot\mathbf{n} +
P^{+}\,\mathbf{n}
\label{gphi2}
\end{equation}
on $\Gamma_c$. Taking the derivative of the Lagrangian with respect to $\Phi$ \Blue{and}
\Blue{taking} into account (\ref{gphi2}) one recovers the gradient
\begin{equation}
\nabla_{\Phi}A_{\mbox{rec}} =\int_{\Gamma_{c}}
 \left(-\frac{1}{Re}\nabla\mathbf{U}^{+}\cdot\mathbf{n} + P^{+}\mathbf{n}\right) \cdot
 \mathbf{B} \; \mbox{d}\mathbf{s},
 \label{sensitivities}
\end{equation}
which relates the modification of the recirculation bubble area
$\delta A_{\mbox{rec}}$
and the modification of the suction 
amplitude 
$\delta \Phi$
through the relation (\ref{sensi}).
%

\Blue{In the following section, we address the design of a control strategy for the baseflow modification. 
The underlying idea is that a (steady) suction induces a baseflow modification and in particular a shortening of the recirculation bubble, which is expected to be qualitatively similar to the action of the unstable dynamics on the time averaged mean flow. The dynamics of the baseflow modification due to suction, that is the transition back to the original steady state (obtained by solving the SFD system (cf. \S 2.2)) will be  characterized in terms of modes applying a Dynamic Mode Decomposition (DMD) which is briefly outlined in Appendix A.} 

\section{Step response \Blue{to a baseflow modification}}


\Blue{We use the hypothesis, that the transition from the modified baseflow back to the steady state is a modal-type dynamics $\delta\mathbf{Q} e^{\sigma t}$, 
with a spatial distribution  $\delta\mathbf{Q}$ (the baseflow  modification) and a decay rate $\sigma$.}
To determine the characteristics of this mode, we consider the step response of the Navier-Stokes equations, that is the dynamics of a modified baseflow with a shorter recirculation bubble, set free to evolve back to the steady state.

However, self-excited instabilities prevent the flow to return to the steady state and the step response is therefore applied to the SFD system (\ref{SFD}) designed to stabilize the Navier-Stokes dynamics.  
We hereafter briefly outline, how the dynamics for the Navier-Stokes system can be inferred from the knowledge about the (stabilized) dynamics of the SFD system.  
As shown by \citet{akervik:brandt:henningson:hoepffner:marxen:schlatter} and \citet{jordi:shervin}, the linear perturbation dynamics evaluated close to a steady state of the Navier-Stokes system (\ref{NSU}) is related to the linear perturbation dynamics of (\ref{SFD}) by a transfer function. 
Indeed, the \Blue{action of the} SFD coupled to the Navier-Stokes equations in the system (\ref{SFD}) \Blue{can be described} by a first order bandpass 
time filter of the form
\citep{akervik:brandt:henningson:hoepffner:marxen:schlatter}
\begin{equation}
\mathbf{R}(\mathbf{x},T) = \int_{-\infty}^{T} \frac 1 \Lambda \exp\left(\frac{\tau-T}{\Lambda}\right) \mathbf{Q}(\mathbf{x},\tau)\,\mbox{d}\tau.
\label{time_filter}
\end{equation}
Performing a Fourier-Laplace transform of the time filter gives
\begin{equation}
\int_{-\infty}^{0} \frac 1 \Lambda \exp\left(\frac{\tau}{\Lambda}\right)\exp(-i\omega\tau)\,\mbox{d}\tau \;=\; \frac{1}{1-i\omega\Lambda},
\end{equation}
where the real part of $\omega$ is the circular frequency. Thus we get the equality for the Laplace transforms 
\begin{equation}
L(\mathbf{R}) = \frac{1}{1-i\omega\Lambda} L(\mathbf{Q}).
\label{TF}
\end{equation}
\Blue{Using this property, the transfer function associated with the SFD system (\ref{SFD}) 
\Blue{provides a linear transformation between}
the stabilized eigenvalues $\sigma_{SFD}$ back to their 
Navier-Stokes
counter-parts $\sigma$.}
Here $-i\omega$ is to be substituted with the stabilized eigenvalue $\sigma_{SFD}$ and introducing (\ref{TF}) in (\ref{SFD}), the dynamics of the
Navier-Stokes baseflow modification without the low-pass filter 
is characterized by the decay rate $\sigma$ through the relation
\begin{equation}\label{eq:SFD_lbd_sfd_trans}
\sigma = \sigma_{SFD} + \zeta\left(1-\frac{1}{1+\sigma_{SFD}\;\Lambda}\right).
\end{equation}
%
\begin{figure}
\begin{center}
\scalebox{0.65}{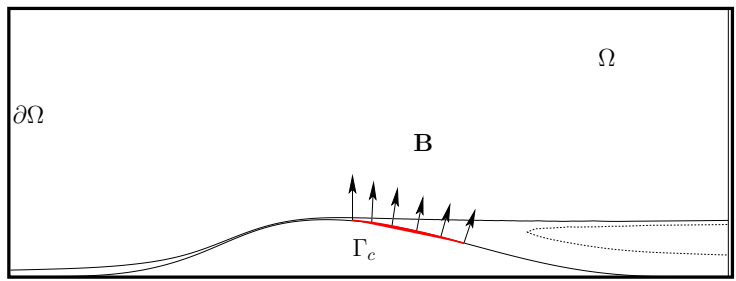}
\caption{Sketch of the domain and of the control setup. {\Blue{
The actuator $\mathbf{B}$ is located on the wall along the surface $\Gamma_c$, located on the boundary $\eta(x) = [25, 30]$.}
}}
\label{bump}
\end{center}
\end{figure}
%

In the following, a baseflow subject to a control is computed as solution of the SFD system (\ref{SFD}). The evolution of the flow state with control back to the original steady state, once the suction suppressed, is then characterized by a dynamic mode decomposition \citep{schmid:2010}, integrating in time system (\ref{SFD}). In this procedure, the baseflow modification decay rate $\sigma_{SFD}$ is recovered and the associated rate for the Navier-Stokes system by the relation (\ref{eq:SFD_lbd_sfd_trans}).

The control is performed by adding a blowing and suction actuator located at the summit of the bump (cf. figure \ref{bump}). \Blue{$\Gamma_c$ denotes the portion of the domain over which the control is applied and was chosen along the boundary $\eta=[25,30]$, which is also the most sensitive region to a forcing. The vector $\mathbf{B}$ is the blowing profile given by the normal $\mathbf{n}$ along $\Gamma_c$} \citep{passaggia:ehrenstein:2013,boujo:ehrenstein:gallaire:2013}.
The evolution of the state with control is obtained by integrating the system (\ref{SFD}) back to the original steady state, once the suction suppressed, is then characterized using
a dynamics mode decomposition.

\subsection{Dynamic mode decomposition of the stabilized step response}\label{sec:DMD_step}

\begin{figure}
\begin{center}
\scalebox{0.85}{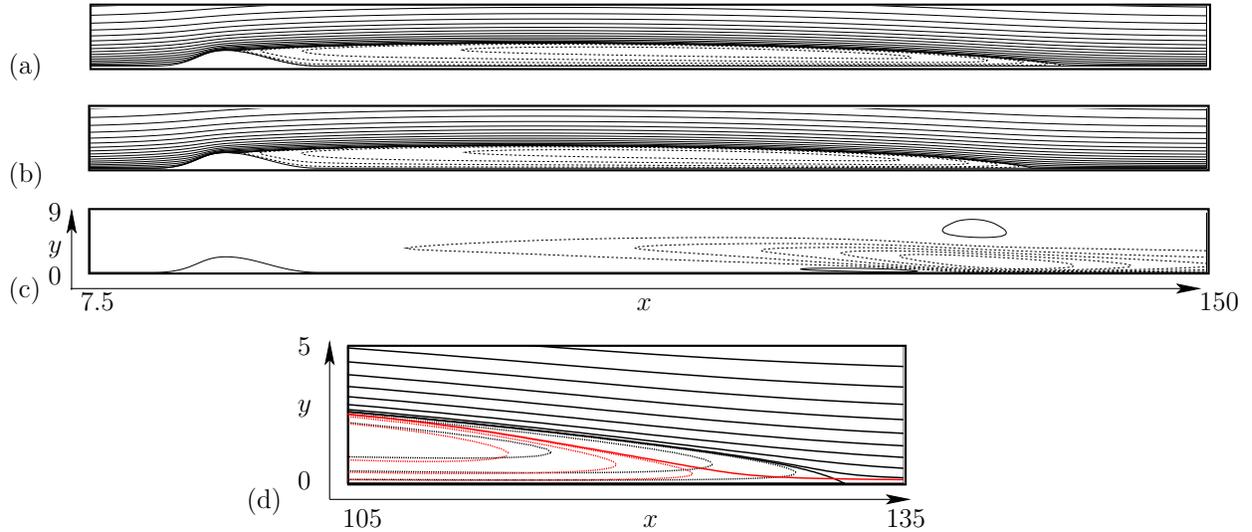}
\caption{Streamlines of the baseflow without (a) and with suction (b) for an 
         amplitude $\Phi_0=-10^{-4}$ imposed by the actuator $\Gamma_c$ at 
         $Re=650$ and a bump height $h=2$. Iso-contours of the streamwise velocity $u$
         component of the baseflow modification (c). The continuous lines denote the 
         positive contours of the associated quantity while the dashed line shows the negative contours.
         Close up of the streamlines (d)
         of the steady state (black) and the suction steady state (red) 
         near the reattachment point, {\Blue{the $y$ axis being stretched compared to the $x$ axis.}}}
\label{suction} 
\end{center}
\end{figure}

\Blue{Evaluating the impulse or step response\footnote{The impulse response of a linear system is the time derivative of the step response, which in the forthcoming dynamic mode decomposition does not affect the frequency content of the response. The DMD analysis of the step response is thus expected to be very close to the impulse response, at least for a small enough step amplitude.} of a linear system can be efficiently performed using either balanced truncations or Krylov-type methods \citep{Antoulas}.}
\Blue{In the case of nonlinear systems, the analogue to Krylov type methods is known as the Dynamic Mode Decomposition (DMD). The DMD method consists in computing a linear approximation of a nonlinear system, solely based on snapshots of the flow \citep{rowley:mezic,schmid:2010}.}
%
%
The initial condition of the step response is a steady state subject to a small amplitude steady \Blue{
suction actuation at the wall chosen as $\Phi_0 = -10^{-4}$.} 
This 
forcing on the stabilized dynamics produces a modified steady state with a slightly 
shorter recirculation bubble (see figures \ref{suction}(a-d)) 
compared with the baseflow.
Note that in \citet{boujo:ehrenstein:gallaire:2013} it has been shown that such steady suction forcing with sufficient amplitude stabilizes the self-sustained instabilities. 
%

The stabilized step response is initialized using the suction steady-state shown in figure \ref{suction}(b). \Blue{Once this actuation is suppressed} (i.e. $\Phi_0=0$) at $t=0$, the stabilized system (\ref{SFD}) is marched in time where the shorter recirculation region evolves
back to its steady state shown in figure \ref{suction}(a).
\begin{figure}
\begin{center}
(a)\scalebox{0.85}{\input{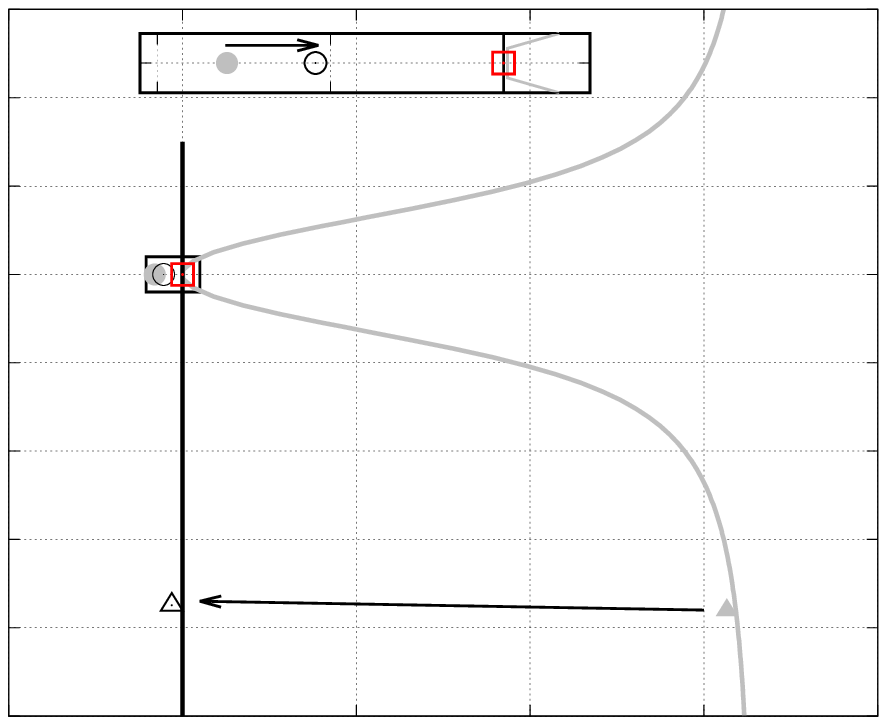}}\\

(b)\hspace{4mm}\scalebox{0.3}{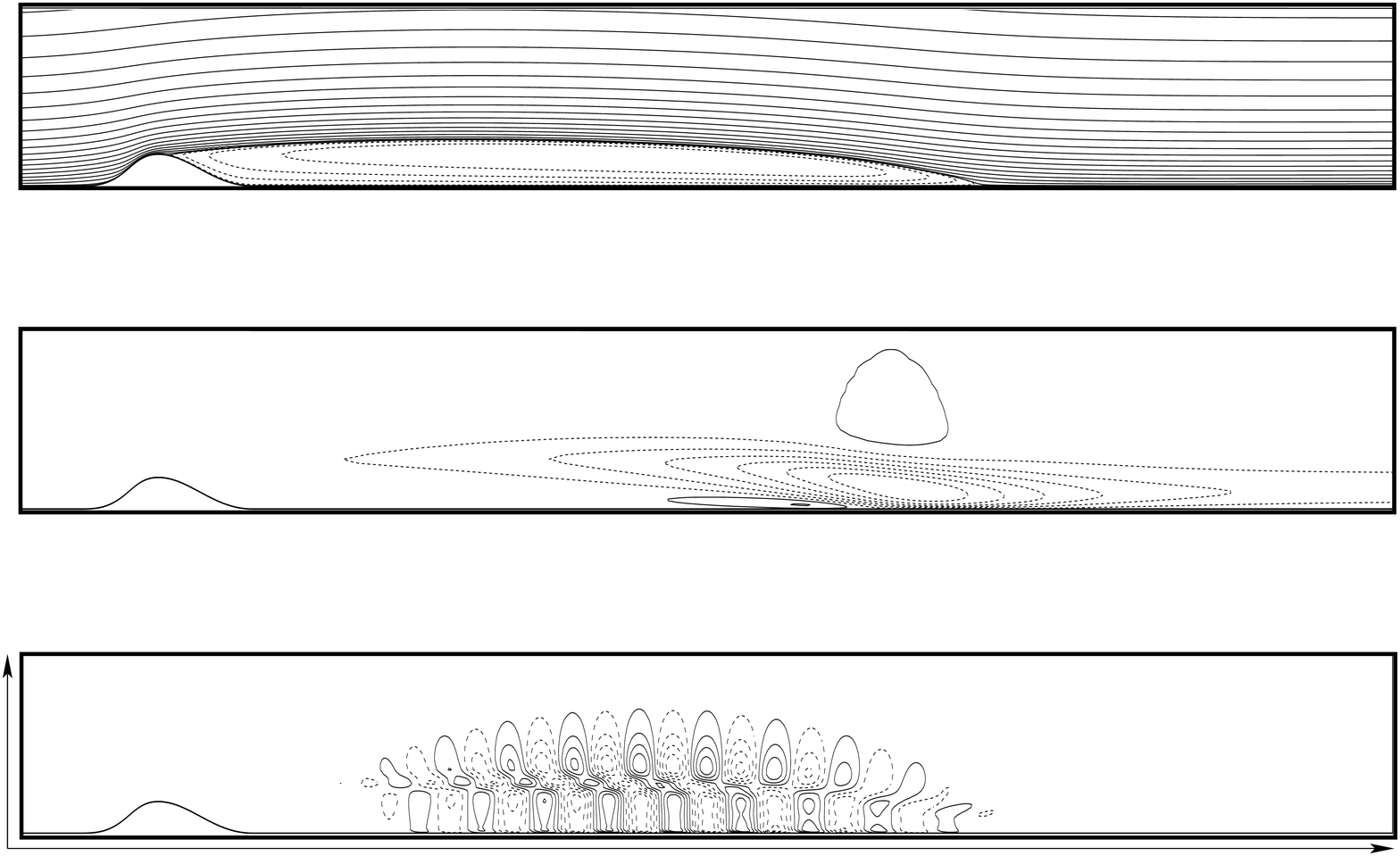}
\caption{(a) DMD spectrum $(\sigma_r , \sigma_i)$ computed by time-marching of the SFD system (empty symbols) and the corresponding Navier-Stokes eigenvalues using the transformation (\ref{eq:SFD_lbd_sfd_trans}) (filled symbols); steady state eigenvalue ($\color{red} \square$), eigenvalue associated with the baseflow modification ($\circ,$ $\; \color{gray} \bullet$), eigenvalue associated with the most amplified oscillatory mode ($\triangle$, $\;\color{gray} \blacktriangle$). The arrows indicate the shift from the Navier-Stokes eigenvalues to the SFD eigenvalues. The black vertical line corresponds to marginal stability ($\sigma_r = 0$) and the corresponding transformed curve using (4.4) is depicted in grey. 
(b) Streamlines associated with the steady state ($\color{red} \square$) (top), iso-contours of the streamwise velocity of the least stable purely real-values mode associated with the baseflow modification (middle) and iso-contours of the {\Blue{streamwise velocity of the most amplified mode $\hat{u}$}} ($\color{gray} \blacktriangle$) (bottom).
}
\label{step_response}
\end{center}
\end{figure}
%
This flow 
evolution 
is analysed using a Dynamic Mode Decomposition (DMD) (see appendix \ref{DMD_A} for a description of the algorithm), 
which will provide 
the temporal mode associated with the baseflow modification.
%

%
%
The DMD analysis has been performed at $\Rey=650$ and the result is shown in figure \ref{step_response}. 
The transformation (\ref{eq:SFD_lbd_sfd_trans}) was applied to the eigenvalues 
$\sigma_{SFD}$ of the DMD spectrum obtained by time-stepping of the SFD system, 
providing the corresponding DMD eigenvalues $\sigma$ for the Navier-Stokes dynamics. 
The least stable real mode is shown in figure \ref{step_response}(a) by mean of a grey circle ($\color{gray} \newmoon$), the grey triangle ($\color{gray} \blacktriangle$) being the eigenvalue associated with the most unstable oscillatory mode, the SFD counterpart being damped as shown by the empty triangle.
The continuous grey line shows the transformation of the imaginary axis through (\ref{eq:SFD_lbd_sfd_trans}).
\citet{cunha:passaggia:lazareff} and \citet{jordi:shervin} showed that the SFD can stabilize the instability modes by decreasing the growth rates of complex eigenvalues when $\zeta$ and 
$\Lambda$ are chosen appropriately. They also showed that the SFD modifies the decay rate of the real eigenvalues $\sigma$ 
associated with the baseflow modification, that is the real eigenvalues $\sigma$ always increase under the action of the SFD, 
independently of the choice of the couple ($\zeta,\Lambda$).
This means that the stabilized decay rate $\sigma_{SFD}$ is always shifted towards the origin when using the SFD method, as shown by the arrow in the close up view in figure \ref{step_response}(a).
%
The steady state is characterized by an eigenvalue equal to zero (i.e. the red square ($\color{red} \square$) in figure \ref{step_response}(a) 
and is shown in figure \ref{step_response}(b).
The real eigenvalue 
(see the grey circle in the close up view near the origin in figure \ref{step_response}(a))
is precisely associated with the baseflow modification shown in figure \ref{step_response}(b), which is seen to be similar 
to the steady state modification $\delta\mathbf{Q}$ shown in figure \ref{etat_base}(d).
\Blue{Therefore, one may assume that the baseflow modification produced by the suction actuator and the baseflow modification induced by the self-excited instabilities are indeed close.}
The most unstable oscillatory mode is also depicted in figure \ref{step_response}(b)
and displays a very similar spatial structure than the most unstable global mode found by \citet{ehrenstein:gallaire:bosse}.

Separating the base flow dynamics from the oscillatory perturbations, we make the assumption of the following decomposition for the unsteady flow field 
\begin{equation}
\mathbf{Q}(\mathbf{x},t) = \mathbf{Q}_0(\mathbf{x}) + \delta\mathbf{Q}(\mathbf{x,t}) + \tilde{\mathbf{q}}(\mathbf{x},t),
\label{base_evol2}
\end{equation}
where $\tilde{\mathbf{q}}$ is the remaining perturbation.
\Blue{Note that by construction, $\tilde{\mathbf{q}}$ satisfies
$$\lim_{\Delta t\to\infty}\frac{1}{\Delta t}\int_{t}^{t+\Delta t} \tilde{\mathbf{q}}(\mathbf{x},t) \;\mbox{d}t=0.$$
In the case of the linear perturbation dynamics, $\delta\mathbf{Q}=0$ and the perturbation is entirely contained in $\tilde{\mathbf{q}}$.} 
Up to now, we have focused on the evolution 
of the baseflow and its modification,
assuming that the fluctuations $\tilde{\mathbf{q}}$ are stabilized (i.e. for values of $\Rey$ below criticality or suppressed by some control), in which case the temporal evolution writes
\begin{equation}
\mathbf{Q}(\mathbf{x},t) = \mathbf{Q}_0(\mathbf{x}) + \delta\mathbf{Q}(\mathbf{x})e^{\sigma t},
\label{base_evol}
\end{equation}
the computed decay rate $\sigma$ associated with the baseflow modification $\delta\mathbf{Q}$ 
being real 
and negative in the present study.
In the next section, a methodology is developed to control the perturbation of the steady state by considering separately the control of the baseflow modification $\delta\mathbf{Q}$ and the fluctuations induced by the instabilities $\tilde{\mathbf{q}}$. This is motivated by the difference of time scales \Blue{shown by mean of DMD eigenvalues in figure \ref{step_response}(a)} between the slow baseflow modification and the \Blue{fast instability modes} such that
\begin{equation}
\partial_t\delta\mathbf{Q} \ll \partial_t\tilde{\mathbf{q}}.
\end{equation}
Using the same actuator, two control strategies are sought separately:
the control of the baseflow modification 
will be optimized using a 
a Linear Quadratic (LQ) type regulator, while the fluctuations will be controlled using an augmented Lagrangian approach of the nonlinear dynamics \Blue{where no assumption is made to compute the control of the flow dynamics}.
\Blue{It is important to note
that it is not necessary to compute $\delta\mathbf{Q}$ explicitly. 
As it will be shown later, 
time derivatives of the flow can indeed be used to filter $\delta\mathbf{Q}$ from the fluctuation $\tilde{\mathbf{q}}$.}
%
\Blue{
Concerning the slow baseflow dynamics, we show in the next section that an analytical expression for the control of the baseflow modification can be computed. This step 
is necessary in order to take into account the slow dynamics when computing the control law. It has again to be emphasized that the baseflow modification 
is precisely the pitfall of the Lagrangian optimization method where finite time of optimization are considered.}

\section{Formulation of the optimization problem}

%

The present control strategy relies on the temporal evolution of a scalar quantity, which is chosen here as the modification of the recirculation area $\delta A_{\mbox{rec}}$, as a measure of the baseflow modification $\delta\mathbf{Q}$ according to (\ref{Arec}). A similar approach was already considered by \citet{boujo:gallaire:2:2014} for the sensitivity analysis of the same bump flow problem and we consider here the temporal evolution of (\ref{base_evol}) subject to a time-dependent control $\Phi$ through a suction actuator located on $\Gamma_c$.
Subsequently to the sensitivity and the step response analyses,  
and based on 
the assumed baseflow dynamics $\delta A_{\mbox{rec}} \; e^{\sigma t}$ (according to (\ref{base_evol})) and the knowledge of the gradient (\ref{sensitivities}) with respect to suction,
a Reduced Order Model (ROM) for the modification of the recirculation
\begin{subeqnarray}
\frac{\mbox{d}\, \delta A_{\mbox{rec}}(t)}{\mbox{d}t} &=& \sigma \delta A_{\mbox{rec}}(t) + \mathcal{G}\Phi(t), \\ 
\Phi(t)&=&K\delta A_{\mbox{rec}}(t)
\label{mean-lin-sys}
\end{subeqnarray}
can be written
\Blue{with $K$ the control gain, which has to be determined, 
and $\mathcal{G}=\nabla_{\Phi}A_{\mbox{rec}}$ the sensitivity.
Note, that according to (\ref{sensi}) for a small steady suction amplitude $\Phi_0$ 
\begin{equation}
\delta A_{\mbox{rec}} = \mathcal{G} \Phi_0.
\label{sensA2}
\end{equation}}


The following subsections provide the main steps of the method.
The results of the sensitivity \Blue{and the step response} analyses are used in \S \ref{sec:control_bf} to design a slowly decaying suction strategy $\Phi(t)$ for the control of the baseflow modification. 
The control of the transients is addressed in \S \ref{sec:control_tr} where a blowing-suction strategy $\phi(t)$ is computed to suppress the instabilities. 

\subsection{Control of the baseflow modification}\label{sec:control_bf}

The aim of the control law $\Phi(t)$ is to
minimize the
\Blue{recirculation area modification $\delta A_{\mbox{rec}}$} and the objective function is
\begin{equation}
\mathcal{J}_{1,int}(\Phi,\delta A_{\mbox{rec}}) = \frac 1 2
 \int_{T_{0}}^{\infty} {\delta A_{\mbox{rec}}}^2(t) \;\mbox{d}t 
+ \gamma \int_{T_{0}}^{\infty} \Phi^2(t) \;\mbox{d}t.
\label{JOL}
\end{equation}
Steady suction near the summit of the bump according to 
the step response analysis and sensitivity analysis of the previous section decreases the recirculation region. 
Here, the aim is to determine a time-dependent control law $\Phi(t)$ that enforces a slow decay of the baseflow modification, slower than the natural decay rate computed in \S \ref{sec:DMD_step}. 
This objective function prevents rapid variations, induced by the baseflow modification itself that can trigger undesired transients, capable of breaking the recirculation region \citep{marxen:henningson}.
This implies that (\ref{JOL}) is to be maximized by conveniently choosing the penalty coefficient $\gamma$.
Therefore we want to find a control law $\Phi(t), t > 0$,
such that the controllable system
(\ref{mean-lin-sys})(a-b) is transferred from an initial state $\delta A_{\mbox{rec}}(t=0)$, to an arbitrary final state 
$\delta A_{\mbox{rec}}(t\rightarrow\infty) = 0$ such that the cost functional (\ref{JOL})
is maximized and attains a finite optimal value. 
The solution of this problem is obtained by solving the algebraic Riccati equation (see appendix \ref{sec:appendix_LQR} for a derivation)
\begin{equation}
-\gamma^{-1} X^2 \mathcal{G}^2 + 2\sigma X + 1 = 0.
\label{ariccati}
\end{equation}
Because 
we aim at maximizing (\ref{JOL}) (i.e. enforce a slower decay rate than the natural decay rate $\sigma$ of the baseflow modification)
for 
$t\rightarrow \infty$, 
we seek the negative 
\Blue{(maximizing)}
solution to the algebraic Riccati equation (\ref{ariccati}), that is 
\begin{equation}
X=\frac{\sigma + \sqrt{\sigma^2 + \gamma^{-1}\mathcal{G}^2}}{\gamma^{-1} \mathcal{G}^2}.
\label{Ricc}
\end{equation}
The control gain $K$ can be computed considering $K=\gamma^{-1}\mathcal{G}X$ (see equation \ref{ansatz2}) and
the time dependence of the recirculation area modification is
the exponential $\exp((\sigma + \mathcal{G} K)t)$.
Finally the control law $\Phi(t)$ is given by
\begin{equation}
\Phi(t)=\Phi_0\exp((\sigma + \mathcal{G} K)t),
\label{control_law}
\end{equation}
for some initial suction forcing $\Phi_0$ 
where $\sigma + \mathcal{G}K \approx -1/\sqrt{\gamma}$, since $\mathcal{G} \gg 1$ 
(typically of order $10^4$) 
and $-\sigma \ll 1$ (of order $10^{-2}$ or even smaller) for the present flow cases.

\subsection{Transients control}\label{sec:control_tr}

For increasing values of the Reynolds number,
transient growth phenomena and global instability 
modes are expected to play a more important role and prevent the flow to return to a steady state.  
Thus transient dynamics control has to be considered and
the Navier-Stokes system is now expressed in perturbation form
$\mathbf{f}(\mathbf{U}_0,\mathbf{q},Re)=0$ with
\begin{equation}
\mathbf{f}(\mathbf{U}_0,\mathbf{q},Re)=[\partial_{t}{\mathbf{u}} +
(\mathbf{U}_0 \cdot \nabla) \mathbf{u} + (\mathbf{u} \cdot \nabla)
\mathbf{U}_0 + (\mathbf{u} \cdot \nabla) \mathbf{u} + \nabla p -
\frac{1}{Re} \nabla^{2} \mathbf{u}; \; \nabla \cdot \mathbf{u}]
\label{linNS}
\end{equation}
where $\mathbf{q}= \mathbf{Q}-\mathbf{Q}_0 = (\mathbf{u},p)^T$ contains the perturbation velocity field 
$\mathbf{u}$ and the pressure $p$ 
and $\mathbf{U}_0$ is the velocity field associated with the baseflow.
It is recalled that this unstable baseflow (at supercritical Reynolds numbers in the absence of suction control) is obtained with the SFD technique and is shown 
in figure \ref{etat_base}(a) for $Re = 650$.

In the following, we seek to compute the time dependent optimal control law $\phi(t)$ that minimizes the cost function $\mathcal{J}_2(\phi(t),\tilde{\mathbf{u}}(t))$ over the finite time window $[T_0, T_1]$. 
Note again that according to the decomposition (\ref{base_evol2}), $\mathbf{q}= \delta\mathbf{Q} + \tilde{\mathbf{q}}$, the distinction between 
the baseflow modification and the transient dynamics being not known a priori. We therefore have to find an alternative to infer the transient perturbation $\tilde{\mathbf{q}}(t)$ based on our knowledge of $\mathbf{q}(t)$.
Minimizing the transient dynamics can be achieved by considering 
the time derivative of the kinetic energy of the perturbation $\mathbf{q}$. 
This is justified as the time derivative of the baseflow modification 
$\delta \mathbf{Q}$ is nearly two orders of magnitudes smaller than 
the time derivative of the transient $\tilde{\mathbf{q}}$. 

Using the time derivative as a high-pass filter of the perturbation dynamics $\mathbf{q}$, 
the objective function is either
\begin{equation}
\mathcal{J}_{2,int}(\mathbf{\phi},\mathbf{u}) =
\frac{1}{2}\int_{T_{0}}^{T_{1}}\int_{\Omega} \frac{\partial\mathbf{u} }{\partial t}\cdot
\frac{\partial\mathbf{u} }{\partial t}\; \mbox{d}\mathbf{x} \,\mbox{d}t + \gamma
\int_{T_{0}}^{T_{1}}\int_{\Gamma_{c}} \mathbf{B}\phi\cdot
\mathbf{B}\phi\; \mbox{d}\mathbf{s}\,\mbox{d}t
\label{jint}
\end{equation}
when energy integral is considered, or
\begin{equation}
\mathcal{J}_{2,term}(\mathbf{\phi},\mathbf{u}) =
\frac{1}{2}\int_{\Omega} \frac{\partial\mathbf{u}(T_{1})}{\partial t}
\cdot \frac{\partial\mathbf{u}(T_{1})}{\partial t}\;
\mbox{d}\mathbf{x} + \gamma
\int_{T_{0}}^{T_{1}}\int_{\Gamma_{c}} \mathbf{B}\phi\cdot
\mathbf{B}\phi\; \mbox{d}\mathbf{s}\,\mbox{d}t
\label{jterm}
\end{equation}
for energy optimization at time $T_{1}$. 
Note that the value of the penalty term $\gamma$ 
is the same as 
in \S\ref{sec:control_bf}.

The control is applied using a blowing-suction signal $\phi(t)$ on $\Gamma_c$ (cf. figure \ref{bump}). 
The boundary condition for the velocity field at $\Gamma_c$ associated with the control $\phi$
has to be added to the Lagrangian as the term
$\int_{T_{0}}^{T_{1}}\int_{\Gamma_{c}} \mathbf{g}(\mathbf{u},\phi)
\cdot \mathbf{B}\phi^+ \;\mbox{d}\mathbf{s}\,\mbox{d}t$, with $\phi^+$ the 
multiplier associated with the control signal $\phi$.

Lagrange multipliers $\mathbf{q}^{+}$ are introduced and the Lagrangian is now
\begin{equation}
\mathcal{L}(\mathbf{q},\mathbf{q}^{+}) = \mathcal{J}_{2} -
 <\mathbf{f}(\mathbf{U}_0,\mathbf{q},Re),\mathbf{q}^{+}> - \; \int_{\Gamma_{c}} \mathbf{g}(\mathbf{u},\phi)\cdot \mathbf{B}\phi^+ \;\mbox{d}\mathbf{s},
\label{lagrangian}
\end{equation}
where $\mathbf{g}(\mathbf{u},\phi)$ is defined as in (\ref{boundcontrol}).
The scalar product $<\cdot>$ is now defined
by the double integral $\int_{T_0}^{T_1}\int_{\Omega} \;\cdot\;
\mbox{d}\mathbf{x}\mbox{d}t$ where the optimization window is taken in
the time interval [$T_0,T_1$]. Similarly to \S 3.1,
the adjoint Navier-Stokes system is obtained by taking the
Fr\'echet derivatives (\ref{Frechet})
of the second term of the Lagrangian (\ref{lagrangian}) \Blue{(see \citet{joslin:gunburger:nicolaides:erlebacher:hussaini,passaggia:ehrenstein:2013} for a complete derivation)}
\Blue{and}
the adjoint system is
\begin{equation}
\begin{split}
\mathbf{f}^{+}(\mathbf{U}_0,&\mathbf{q},\mathbf{q}^{+},Re)= \bigg{[}
-\partial_{t}{\mathbf{u}^{+}} -(\mathbf{U}_0 \cdot \nabla)
\mathbf{u}^{+} + (\nabla\mathbf{U}_0)^{T} \mathbf{u}^{+}\\ -&
(\mathbf{u} \cdot \nabla) \mathbf{u}^{+} + (\nabla\mathbf{u})^{T}
\mathbf{u}^{+} + \nabla p^{+} - \frac{1}{Re} {\nabla}^{2}
\mathbf{u}^{+} ; - \nabla\cdot\mathbf{u}^{+}\bigg{]}.
\end{split}
\label{adjoint}
\end{equation}
The boundary terms 
are now given by
\begin{equation}
\begin{split}
& \int_{T_{0}}^{T_{1}}\int_{\partial\Omega} \bigg{[}
\frac{1}{Re}\mathbf{u}^{+}(\nabla\hat{\mathbf{u}}\cdot \mathbf{n})
-\frac{1}{Re}(\nabla\mathbf{u}^{+}\cdot \mathbf{n}) \hat{\mathbf{u}}
-(\mathbf{U}_0\cdot \mathbf{n})\mathbf{u}^{+}\cdot \hat{\mathbf{u}}
-(\mathbf{u}\cdot \mathbf{n})\mathbf{u}^{+}\cdot \hat{\mathbf{u}}\\
&+(p^{+}\cdot\mathbf{n})\hat{\mathbf{u}} - \mathbf{u}^{+}(\hat{p}\cdot
\mathbf{n}) \bigg{]} \; \mbox{d}\mathbf{s} \;\mbox{d}t \; + \;
\int_{\Omega} [\mathbf{u}^{+}\cdot\hat{\mathbf{u}}]_{T_0}^{T_1}\;
\mbox{d}\mathbf{x},
\end{split}
\label{bord}
\end{equation}
\Blue{and}the boundary conditions for the perturbation velocity field $\mathbf{u}$ are 
similar to the baseflow
\begin{equation} 
\frac{1}{Re} (\nabla \mathbf{u}) \cdot 
\mathbf{n} \Big{|}_{x=L} = p \mathbf{n} \Big{|}_{x=L}, 
\quad \mathbf{u}\Big{|}_{x=0} = 0,
\quad \mathbf{u}\Big{|}_{y=\eta,H} = 0,
\label{bounddirect}
\end{equation}
but the boundary integral in (\ref{bord}) now vanishes when using
\begin{equation}
 \frac{1}{Re}(\nabla\mathbf{u}^{+})\mathbf{n}
+ (\mathbf{U}_0\cdot \mathbf{n})\mathbf{u}^{+}
+ (\mathbf{u}\cdot \mathbf{n})\mathbf{u}^{+}\Big{|}_{x=0,L}
= p^{+}\mathbf{n}\Big{|}_{x=0,L}, \quad\quad\\
\mathbf{u}^{+}\Big{|}_{y=\eta,H}=0.
\label{boundadjoint}
\end{equation}
The boundary condition (\ref{boundadjoint}) imposed at the inlet
allows
the adjoint solution, which has to be integrated backward in time 
from $T_1$ to $T_0$, to leave the domain without reflections 
and the inlet may therefore be chosen  
relatively close to the bump.
The boundary conditions (\ref{boundadjoint}) are used \Blue{in addition to}  
$\mathbf{u}^+|_{\Gamma_c} = 0$. There are remaining terms in the boundary 
integral (\ref{bord}) and 
\Blue{similarly to equation (\ref{gphi2})}
\Blue{these terms cancel when}
\begin{equation}
\mathbf{B}\phi^{+} = -\frac{1}{Re}\nabla\mathbf{u}^{+}\cdot\mathbf{n} +
p^{+}\,\mathbf{n}
\label{gphi}
\end{equation}
on $\Gamma_c$. 
%
Taking into account (\ref{gphi}), the gradient
\begin{equation}
\nabla_{\mathbf{\phi}}\mathcal{J}_{2}({\phi}) =
 \gamma\phi\int_{\Gamma_{c}}\mathbf{B}\cdot \mathbf{B}
 \;\mbox{d}\mathbf{s} +\int_{\Gamma_{c}}
 \left(-\frac{1}{Re}\nabla\mathbf{u}^{+}\mathbf{n} + p^{+}\mathbf{n}\right) \cdot
 \mathbf{B} \; \mbox{d}\mathbf{s},
\label{gradient2}
 \end{equation}
\Blue{is obtained by considering the Fr\'echet derivative of the Lagrangian with respect to $\phi$ and equation (\ref{gradient2})} is used in a line search algorithm to minimize either (\ref{jint}) or (\ref{jterm}).

%
\Blue{The Fr\'echet derivative of the cost function (\ref{jint}) with respect to $\mathbf{q}$ generates the extra term 
$\int_{T_{0}}^{T_{1}}\int_{\Omega}\partial_t\mathbf{u}\cdot \partial_t\hat{\mathbf{u}}\;\mbox{d}\mathbf{x}\,\mbox{d}t$.}
In that case the adjoint system (\ref{adjoint}) 
is subject to a source term, that is
\begin{equation}
\mathbf{f}^{+}(\mathbf{U}_0,\mathbf{q},\mathbf{q}^{+},Re) = \left[ \frac{\partial\mathbf{u}}{\partial t} ; 
0 \right], \quad \mbox{with} \quad \mathbf{u}^+ (T_1) = 0.
\label{source}
\end{equation}  
Note that similar to the procedure used to solve (\ref{NSU}), the system 
$\mathbf{f}(\mathbf{U}_0,\mathbf{q},Re) = 0$ for the 
flow perturbation with the boundary conditions (\ref{bounddirect}) 
and the adjoint system (\ref{source}) 
subject to the conditions (\ref{boundadjoint}) are solved using 
again the influence matrix technique. The 
direct flow velocity field $\mathbf{u}$ is obtained through 
time-marching from $T_0$ and $T_1$ and it enters the adjoint system 
which is to be solved backward in time from $T_1$ to $T_0$.\\

The difference between the energy time-integral optimization and the 
optimization at $T_1$ is that in the latter case, instead of 
(\ref{source}), the homogeneous adjoint system is solved for a 
non-zero initial condition at $T_1$, that is
\begin{equation}
\mathbf{f}^{+}(\mathbf{U}_0,\mathbf{q},\mathbf{q}^{+},Re) = 0 \quad 
\mbox{with} \quad \mathbf{u}^+ (T_1) = \frac{\partial\mathbf{u} (T_1)}{\partial t}.
\end{equation}
It has again to be emphasized that the perturbation $\mathbf{q}$ includes the slowly varying baseflow modification and the expected transient control signal $\phi$ superimposes to the control law
$\Phi(t)$ given by (\ref{control_law}).

\section{Results}

The control of the present geometry was already investigated in \citet{ehrenstein:passaggia:gallaire} using model reduction and in \citet{passaggia:ehrenstein:2013} using the augmented Lagrangian approach where both methods were not capable to control the nonlinear dynamics, even for Reynolds number close to criticality.
\citet{passaggia:ehrenstein:2013} showed that
the control performances of the augmented Lagrangian approach were degrading for increasing Reynolds numbers 
and the flow seemed to be nearly uncontrollable for $\Rey>650$. In the following, we assess if the suction 
strategy (\ref{control_law}) is sufficiently robust with increasing Reynolds number and how the control law 
can be further optimized. We also want to assess if the present strategy 
allows of driving 
the flow close enough to its steady state where reduced-order type feedback controllers, capable 
of controlling the linear perturbation dynamics, could be coupled.

\begin{figure}
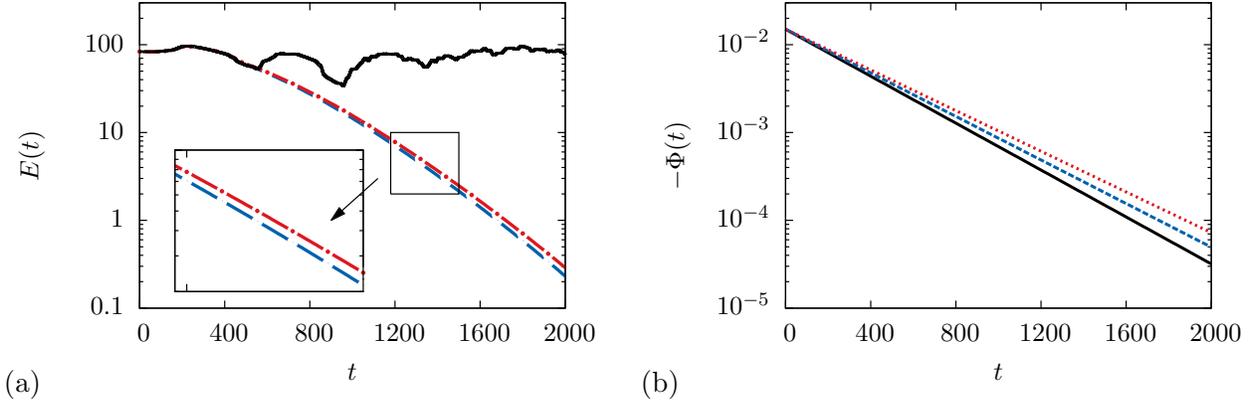

\begin{center}
\begin{minipage}[t]{0.49\textwidth}
(a)\hspace{-4mm}\scalebox{0.9}{\input{Energie_610.tex}}
\end{minipage}
\begin{minipage}[t]{0.49\textwidth}
(b)\hspace{-3mm}\scalebox{0.9}{\input{Phi_610.tex}}
\end{minipage}
\caption{Control of 
the nonlinear dynamics at $Re = 610$. 
Energy of the perturbation $E(t)=\frac 1 2\int_\Omega \mathbf{u}^2\mbox{d}\Omega$ for the controlled dynamics at $\gamma=1.06\,10^5$ (------), 
$\gamma=1.22\,10^5$ $(\textcolor{blue}{- - -})$, $\gamma=1.41\,10^5$ $(\textcolor{red}{-\cdot-\cdot-})$ using $\mathcal{J}_{1}$ (a). Associated Control laws (b) with $\Phi_0=-1.5\, 10^{-2}$ and where 
the decay rate $\sigma+\mathcal{G}K=$ $-3.10\,10^{-3}$ (-----), $-2.86\,10^{-3}$ $(\textcolor{blue}{- - -})$ 
and $-2.67\,10^{-3}$ $(\textcolor{red}{\cdots})$.}
\label{control_nl_610}
\end{center}
\end{figure}

The control law (\ref{control_law}) has been used as an open-loop strategy for different 
values of $\gamma$ in figure \ref{control_nl_610}(a,b) at $\Rey=610$.
This flow configuration is characterized by very weakly unstable global modes \citep{ehrenstein:gallaire:bosse}.
When subject to a continuous suction forcing $\Phi_0=-1.5\, 10^{-2}$, the area of the recirculation region at steady state $A_{\mbox{rec}}=230.4$ decreases and the flow becomes globally stable
\citep{boujo:ehrenstein:gallaire:2013}.
At this Reynolds number, the decay rate associated with the baseflow modification is $\sigma\approx -0.02$. The sensitivity prior to a suction actuator located at  the summit of the bump (i.e. $x=25$) is ${\cal G}=1.078\times10^4$.
The control strategy (\ref{control_law}) is integrated in time for different values of the control cost $\gamma$. 
Note that increasing $\gamma$ attenuates the decay rate $\sigma + \mathcal{G}K$ of the controlled baseflow modification dynamics.
The evolution of the energy is shown in figure \ref{control_nl_610}(a). During the early times of the simulations, the energy first increases when the suction starts to decrease. This is due to a single vortex, shed from the recirculation region which is rapidly advected downstream. When the control decrease is sufficiently slow, the decay of the energy decreases nearly exponentially in time, according to the control decay rate imposed by the actuator and shown in figure \ref{control_nl_610}(b).
However, for $\gamma < 1.05\times10^5$, that is for lower control costs, the control is not capable to stabilize the dynamics and after $600$ time units, the flow returns to a saturated nonlinear state (cf. figure \ref{control_nl_610}(a)). 

\begin{figure}
\begin{center}
\begin{minipage}[t]{0.49\textwidth}
(a)\hspace{-6mm}\scalebox{0.84}{\input{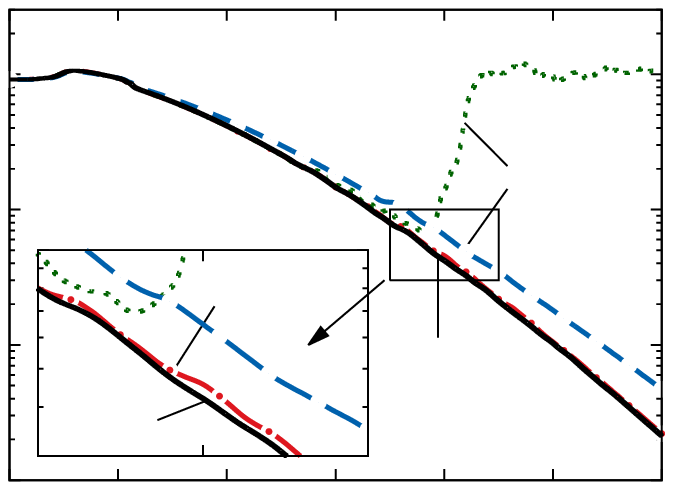}}
\end{minipage}
\begin{minipage}[t]{0.49\textwidth}
(b)\hspace{-2mm}\scalebox{0.87}{\input{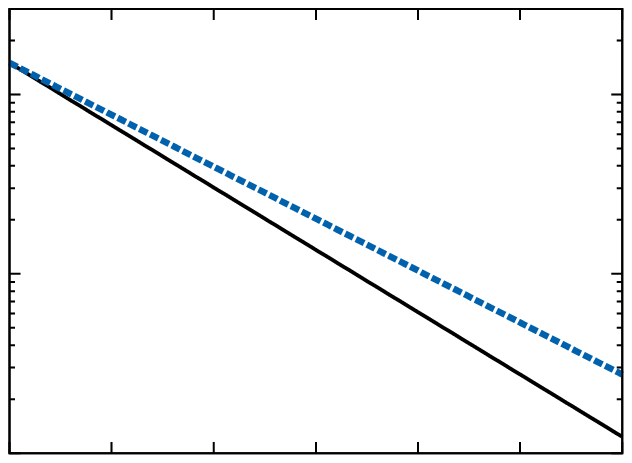}}
\end{minipage}
\begin{minipage}[t]{0.49\textwidth}
(c)\hspace{-4mm}\scalebox{0.9}{\input{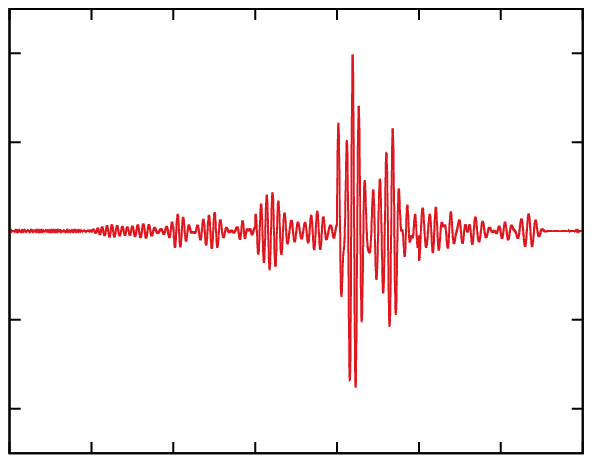}}
\end{minipage}
\begin{minipage}[t]{0.49\textwidth}
(d)\hspace{-3mm}\scalebox{0.9}{\input{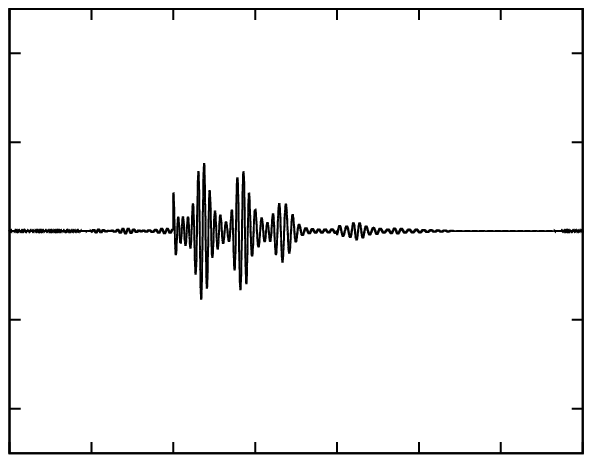}}
\end{minipage}
\caption{Control of the nonlinear dynamics at $Re = 650$.  
Energy of the perturbation for the open loop control strategy at $\gamma=2.5\times10^{5}$ $(\textcolor{green}{\cdots})$,
with additional 
optimized dynamics using $\mathcal{J}_{2,term}$ $(\textcolor{red}{-\cdot-\cdot-})$ 
and using $\mathcal{J}_{2,int}$ (-----); open loop control for $\gamma=3.6\times10^{5}$ $(\textcolor{blue}{- - -})$ (a).
Optimal control laws $\Phi(t)+\phi(t)$ computed with $\mathcal{J}_{2,term}$ (b) at $\gamma=2.5\times10^{5}$ (-----) and  $\gamma=3.6\times10^{5}$ $(\textcolor{blue}{- - -})$; for $\gamma=2.5\times10^{5}$, 
$\phi(t)$ computed with $\mathcal{J}_{2,term}$ (c) and 
$\phi(t)$ computed with $\mathcal{J}_{2,int}$ (d).}
\label{control_nl_650}
\end{center}
\end{figure}

\begin{figure}
\begin{center}
\scalebox{0.5}{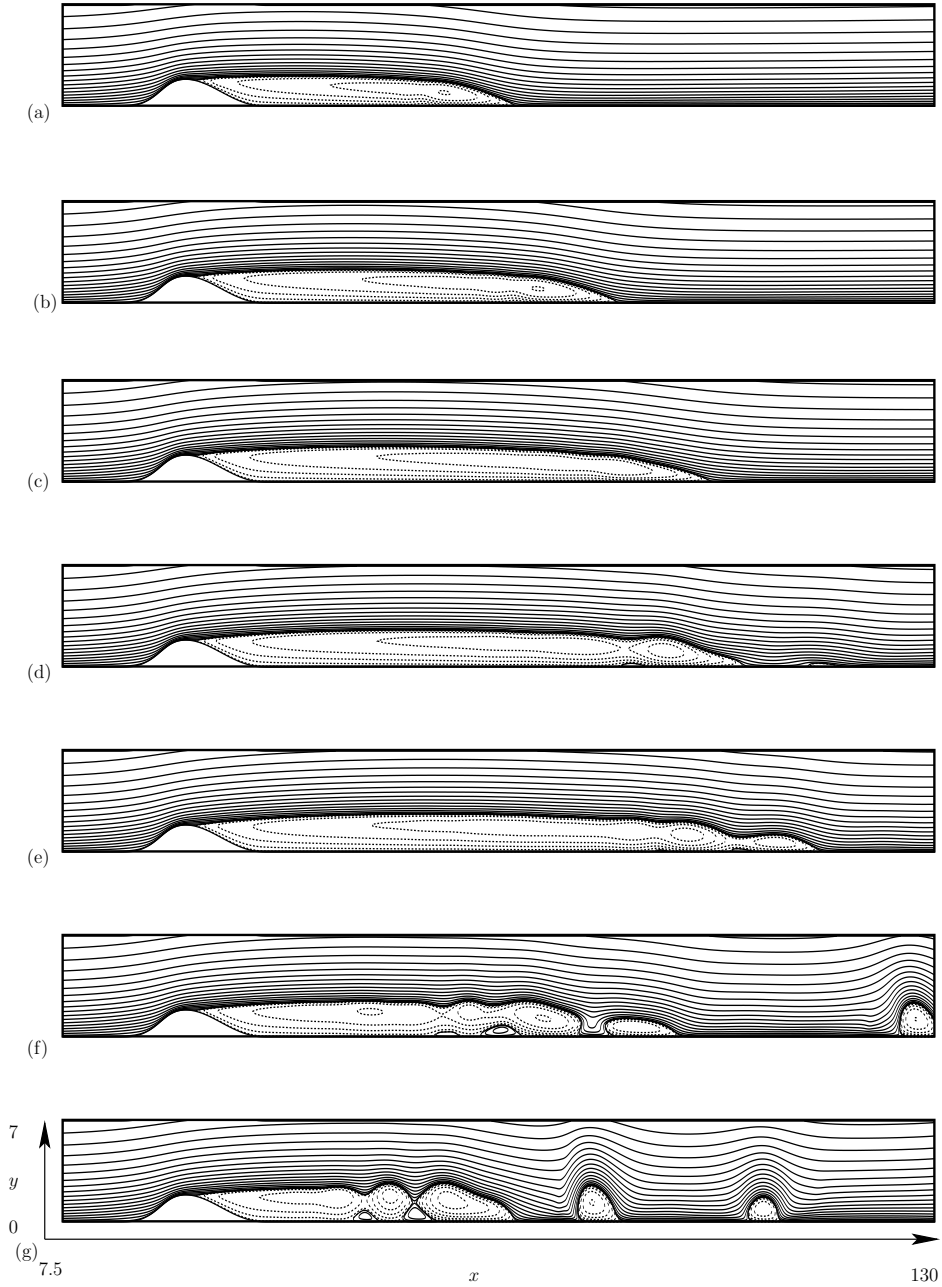}
\caption{{\Blue{Sequence of streamlines for the failing open-loop strategy for $\gamma=2.5\times10^5$ and $\Rey=650$ at (a) $T=300$, (b) $T=600$, (c) $T=900$, (d) $T=1200$, (e) $T=1500$, (f) $T=1800$, and (g) $T=2100$ for the terminal time control strategy, using the objective function $\mathcal{J}_{2,term}$. Note that the $y$ axis is stretched compared to the $x$ axis.}}}
\label{seq_fail}
\end{center}
\end{figure}

The control of the fully nonlinear flow state is now considered using the suction control
strategy described in \S\ref{sec:control_bf} at $\Rey=650$. 
The recirculation area of the steady state is $A_{\mbox{rec}}=246$, the sensitivity analysis provides 
the value $\mathcal{G}=1.49\times10^4$, the decay rate being $\sigma\approx -0.003$ and the initial suction control $\Phi_0=-1.5\times10^{-2}$ proved suitable for reaching a steady state. 
\Blue{The time evolution of the energy shown in figure \ref{control_nl_650}(a) produces similar dynamics to what was previously observed in the $\Rey=610$ case.}
As shown in figure \ref{control_nl_650}(a), a value of $\gamma=3.6\times10^5$ and thus 
a time-dependent decreasing suction of the form (\ref{control_law}) with 
$\sigma+\mathcal{G}K=-3.3\times10^{-3}$
forces the flow to return to a steady state. Decreasing $\gamma$ to $\gamma=2.5\times10^5$ and thus $\sigma+\mathcal{G}K=-2.\times10^{-3}$ 
(cf. figure \ref{control_nl_650}(b)) appears to offer a faster control but at $t\approx1500$,
transient growth followed by the low-frequency flapping instability makes the flow to return to the saturated nonlinear regime \Blue{(see figure \ref{seq_fail} for a sequence illustrating this process)}.
\Blue{This failing strategy is characterized by a large amount of transient growth at $t\approx1500$ and was studied for similar flow conditions by \cite{ehrenstein:gallaire:bosse}.}

To prevent the flow to transit back to the perturbed state (at $\gamma = 2.5 \times 10^5$), the strategy described in 
\S\ref{sec:control_tr} has been applied during the time-marching in order to suppress the transients.
The control signal $\phi(t)$ is solved for the time of optimization $[T_0,T_1]$, where $T_1-T_0=600$, until convergence of the gradient between two iterates 
with a residuum 
$10^{-3}$ is achieved. Note that in \citet{passaggia:ehrenstein:2013}, the convergence
of the gradient appeared to be difficult to achieve and the gradient (\ref{gradient2}) had to be evaluated $10$ to $15$ times for each time window, in order to find a local minimum of their cost function.  In the present investigations, only two to three 
evaluations of (\ref{gradient2}) for each optimization time window proved to be necessary to decrease (\ref{jint}) or (\ref{jterm}) as well as (\ref{gradient2}) by two orders of magnitude and converge to a local minimum of the cost functions. 
The control is then restarted for a time shifted by half the time of optimization ($T_{a}=(T_1-T_0)/2=300$) and marched for a new time interval $[T_0+T_{a},T_1 +T_{a}]$.
Note that \citet{passaggia:ehrenstein:2013} considered an objective function based on the energy, whereas here the square of the time derivative of the velocity is considered. Also, the initial state they departed from was a fully developed nonlinear state associated with the vortex shedding showed in figure \ref{etat_base}(c). In the present study, the suction state from which we start controlling the flow does not exhibit vortex shedding, and proves to be more convenient to control the flow back to a steady state.
%
%
%

\begin{figure}
\begin{center}
\scalebox{0.65}{\input{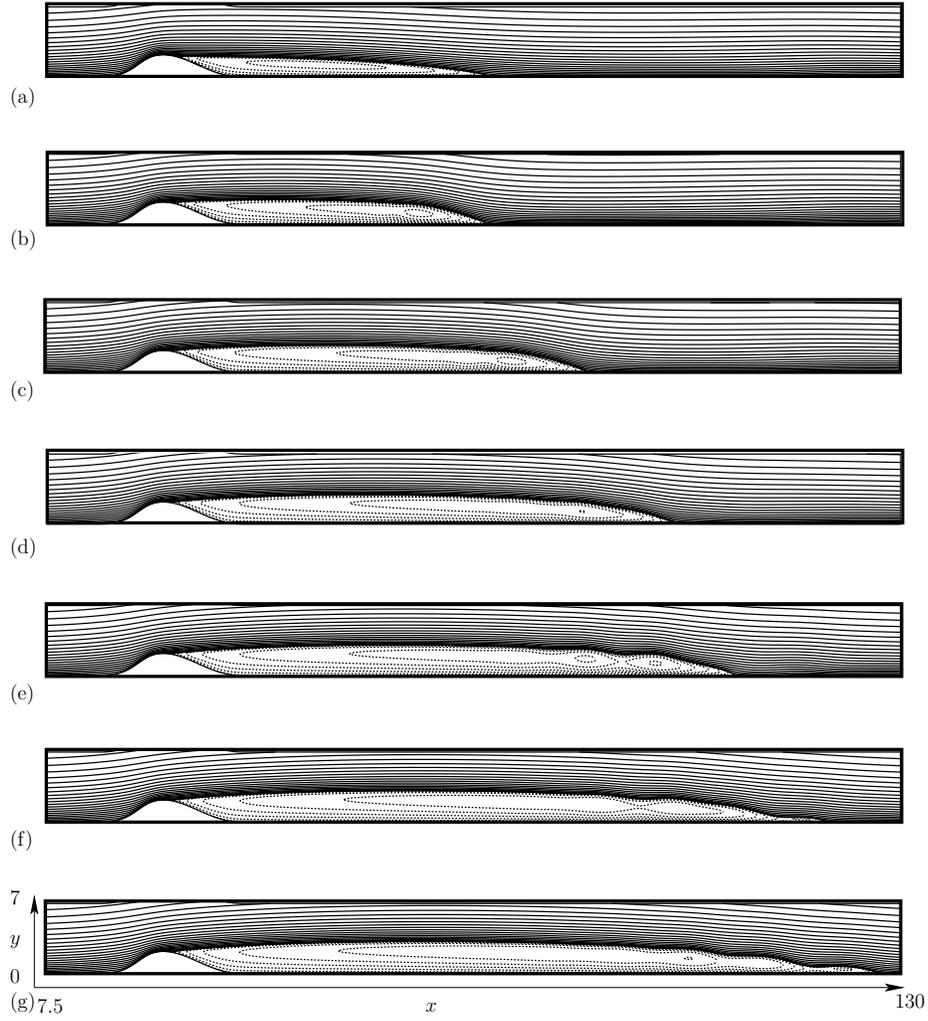}}
\caption{Sequence of streamlines for the adjoint-based controlled strategy for $\gamma=2.5\times10^5$ and $\Rey=650$ (a) $T=0$, (b) $T=300$, (c) $T=600$, (d) $T=900$, (e) $T=1200$, (f) $T=1500$, and (g) $T=1800$ for the terminal time control strategy, using the objective function 
$\mathcal{J}_{2,term}$. Note that the $y$ axis is stretched compared to the $x$ axis.}
\label{control_seq}
\end{center}
\end{figure}

\begin{figure}
\begin{center}
\scalebox{0.5}{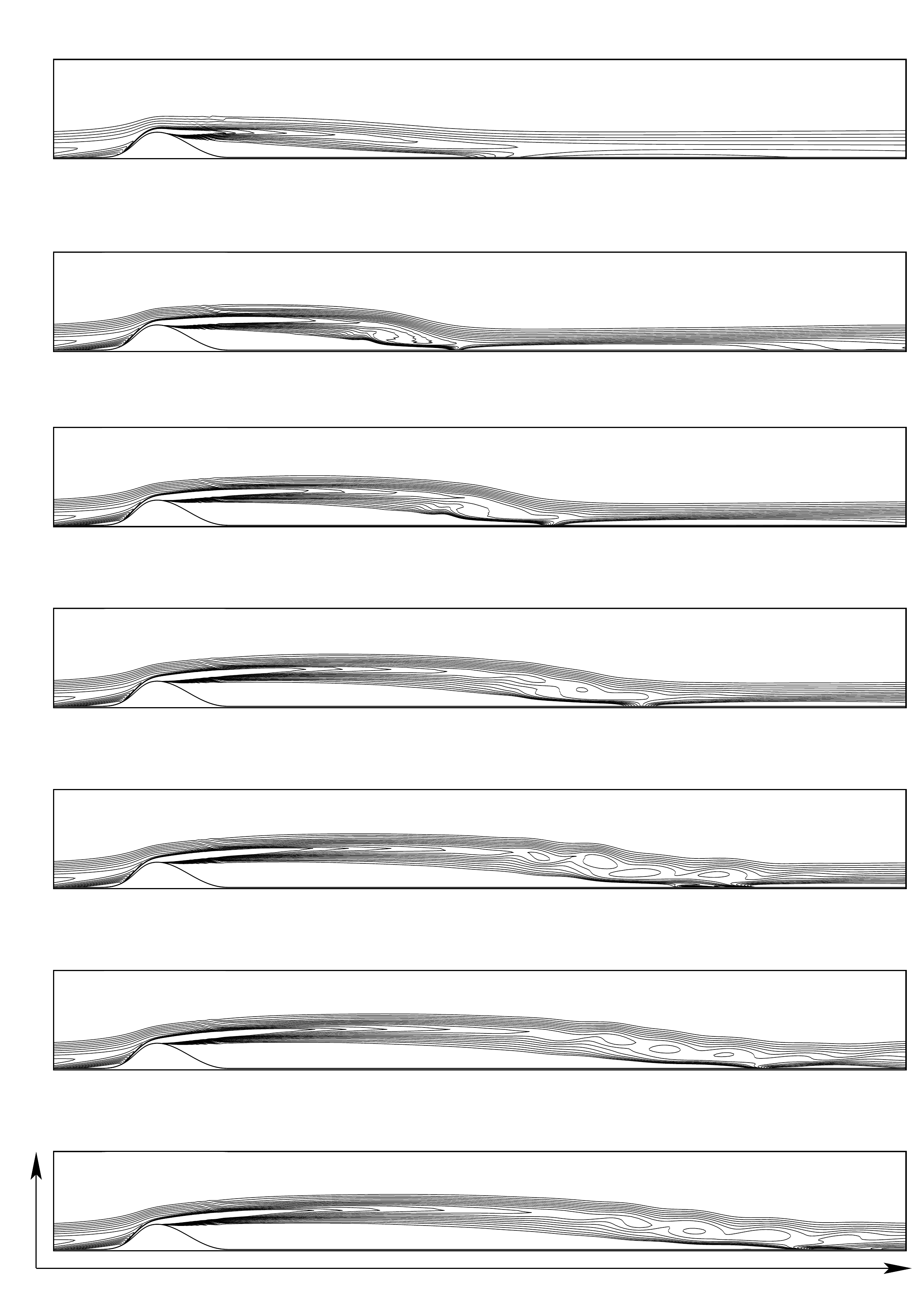}
\caption{Same as figure \ref{control_seq} showing the time evolution of arbitrary contours of positive vorticity. {\Blue{Note that the $y$ axis is stretched compared to the $x$ axis.}}}
\label{seq_vort}
\end{center}
\end{figure}

Both control objective functions $\mathcal{J}_{2,term}$ and $\mathcal{J}_{2,int}$ 
proved suitable 
for controlling the 
low-frequency flapping-type instability 
(typical for this recirculation bubble as discussed in \cite{ehrenstein:gallaire:bosse}),
while the flow reaches its steady state.
A closer look at the evolution of the energy in figure \ref{control_nl_650}(a) shows that $\mathcal{J}_{2,int}$ 
performs
however a little 
better than $\mathcal{J}_{2,term}$.
%
The control laws for both objective function $\mathcal{J}_{2}$ are shown in figures \ref{control_nl_650}(c-d). The terminal time control has a larger amplitude and acts later than the integral time control.
The latter strategy essentially controls the transients early in the simulation, in the range $t=[600,1100]$ whereas the terminal 
is less involved in the beginning but compensates later at $t=[1200,1800]$. Both control strategies $\phi(t)$ are characterized by 
two frequencies. Performing a fast Fourier transform of $\phi(t)$, a high frequency $f_{high}\approx4.67\times10^{-2}$ associated with the unstable global modes and
$f_{low}\approx4.76\times10^{-3}$ corresponding to the low frequency instability dominate,
the different instability characteristics being reported in  \citet{passaggia:leweke:ehrenstein} and \citet{ehrenstein:gallaire:bosse} for 
this bump-induced separated boundary-layer flow.
Note that the amplitude of the control for the transients $\phi(t)$ is $5$ to $6$ orders of magnitudes smaller 
than the suction strategy $\Phi(t)$ and is therefore not noticeable in figure \ref{control_nl_650}(b).

The time evolution of the flow, computed for the objective function $\mathcal{J}_{2,term}$ is shown in figure \ref{control_seq}. Starting from the modified steady state, subject to the steady suction $\Phi_0$, the recirculation region extends slowly towards its original length. As pointed out earlier, the effect of the low-frequency flapping instability starts to be visible at $t=1200$ together with a very weak vortex shedding observed at $t=1200$. At time $t=2400$, the flow is completely stabilized and the flow resembles the steady state, shown in figure \ref{etat_base}(a). 
\Blue{The effect of the control can be observed in figure \ref{seq_vort}(e-g) by mean of iso-contrours of vorticity where the shear layer displays a train of vortices, upstream the reattachment region. This process is the result of the wave-canceling phenomena induced by the control \cite{joslin:gunburger:nicolaides:erlebacher:hussaini} where small amplitude wave packets, excited upstream by the actuator, are amplified in the shear layer and stabilize the flapping instability.}
Note that for $\gamma<2.2\times10^5$ the suction control fails at stabilizing the unstable dynamics.

\begin{figure}
\begin{center}
\begin{minipage}[t]{0.49\textwidth}
(a)\hspace{-3mm}\scalebox{0.9}{\input{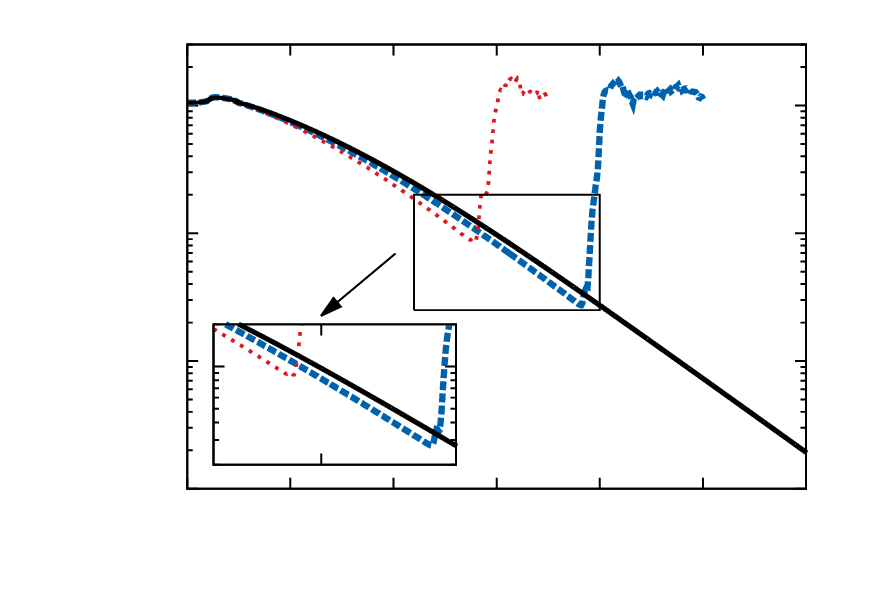}}
\end{minipage}
\begin{minipage}[t]{0.49\textwidth}
(b)\hspace{-2mm}\scalebox{0.9}{\input{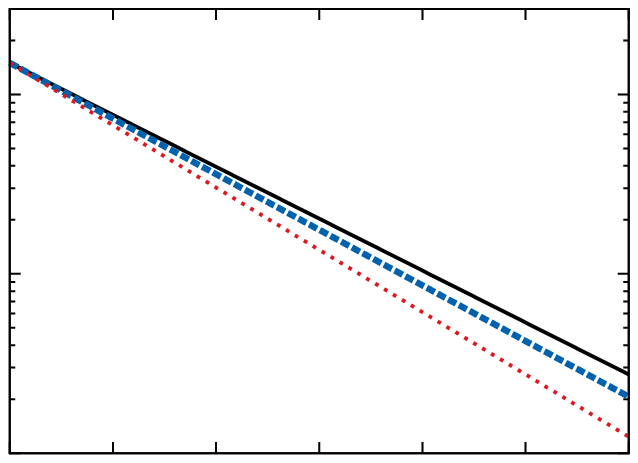}}
\end{minipage}
\caption{Control of the nonlinear dynamics at $\Rey = 700$. 
Energy of the perturbation for the suction strategy at $\gamma=2.25\times10^{6}$ (-----), $\gamma=1.96\times10^{6}$ $(\textcolor{blue}{- - -})$ and
$\gamma=1.56.\times10^{6}$ $(\textcolor{red}{\cdots})$ (a). Associated optimal control laws $\Phi(t)$ (b).}
\label{control_nl_700}
\end{center}
\end{figure}

The same procedure is used for a more unstable case at $\Rey=700$. 
For this highly supercritical flow regime, the recirculation area of the steady state increases to $A_{\mbox{rec}}=266.2$, the sensitivity analysis provides the value $\mathcal{G}=1.255\times10^4$ 
and the initial suction control is again $\Phi_0=-1.5\times10^{-2}$. Note that in that case, 
the absolute value of the decay rate of the baseflow modification decreases to 
$ - \sigma = 0.00115$ which is 
much smaller 
than the value $- \sigma \approx 0.004$ found at $\Rey=650$ and  
the control law $\Phi(t)$ is expected to behave accordingly.  
This is observed in figure \ref{control_nl_700}(a), a value of $\gamma=2.25\times10^{6}$ and thus $\sigma + \mathcal{G}K=-6.67\times10^{-4}$
(cf. figure \ref{control_nl_700}(b)) allowing of controlling back to a steady state. Attempts to control
the transients associated with faster control strategies (i.e. $\gamma=1.56\times10^{6}$ and $\gamma=1.96\times10^{6}$) were made. However at $t\approx2600$ or $t\approx4000$, the transients followed by the low frequency flapping instability appear to be uncontrollable using the present optimization algorithm. 

The transients associated with this phenomena
trigger a wave packet near the separation point, whose transient energy growth appears to be comparable to the optimal perturbation,
which is likely to be uncontrollable using a single actuator \citep{ehrenstein:passaggia:gallaire,passaggia:ehrenstein:2013}. This finite amplitude wave packet is sensitive to the decay rate of the control law $\Phi(t)$ (cf. figure \ref{control_nl_700}), that is when controlled too fast, the flow triggers transient growth. This phenomena forces the flow to return to the saturated nonlinear regime, despite any attempts to control the flow dynamics.
As can be seen in figure \ref{control_nl_700}, the more expensive (in terms of suction amplitude) control with $\gamma=2.25\times10^{6}$ leads to a continuous 
decrease of the perturbation energy and no transient dynamics has been observed up to $t = 8000$ using 
this suction strategy. We did not attempt to progress further in time, but the baseflow modification will ultimately become negligible. As the baseflow approaches the steady state, the self-excited instabilities associated with the resonator dynamics \citep{ehrenstein:gallaire:bosse,passaggia:leweke:ehrenstein} are expected to grow again. However, in this quasi-linear regime, \citep{passaggia:ehrenstein:2013} already performed the control of the self-excited instabilities successfully by computing $\phi(t)$ using the direct-adjoint optimization procedure.
%
%
%

%
%

\section{Concluding Discussion}
A new control methodology has been proposed, using sensitivity analysis associated with baseflow modification, optimal baseflow control and Lagrangian-based optimization procedures for transient perturbation dynamics.
As a flow example a separated boundary layer has been considered and it is shown, 
that when starting the simulation far from the steady state, it is possible to control this globally unstable flow using a time vanishing suction strategy at the wall, at least for sufficiently slow exponential suction decay rates. 
The present strategy 
takes advantage of 
the high sensitivity of the baseflow modification to suppress the instabilities and successfully drive the flow back to its steady state.
The present approach to control the baseflow modification leads to a simple expression for the slowly decaying suction control that could be easily implemented in an experiment. 
The baseflow modification could for instance be inferred from the difference between a local measurement of the flow and numerically computed steady state data.  
This suction strategy to control the baseflow can be further optimized when aiming at a lower control cost, using a Lagrangian-based optimization capable of suppressing transient dynamics.

A recent study showed that the augmented Lagrangian approach could 
for instance be used to control the flow behind a cylinder \citep{flinois:colonius} for values of the Reynolds number well beyond criticality. However the method requires very long time horizons for the optimization, which 
becomes numerically intractable for complex flow dynamics or three-dimensional configurations.
The present approach, combining baseflow modification control and optimization using the augmented Lagrangian approach, could be a possibility to reduce the time horizons, 
which is the limiting factor in the case of complex three-dimensional applications for flow optimization algorithms.

The decay rate at which the flow can be controlled appears to be essentially limited by the effects of transient growth. 
A more robust approach would consist in redesigning the control and assign the long time suction strategy to asymptotically reach a finite value, associated with a more stable steady state. Also, more actuators could be implemented inside the recirculation region or upstream the bump, which could provide more leverage on the control of the baseflow modification.

Finite amplitude perturbations or external noise disturbances
would also be a limiting factor,
when considering real separated flow cases at high Reynolds numbers. 
In particular, the control of incoming boundary-layer instabilities upstream the
separation region 
is likely to be mandatory, in order to decrease or suppress the noise amplified by the boundary layer.
Control methods for boundary-layer instabilities in the absence of a pressure gradient have already been proposed and could be implemented 
\citep{joslin:gunburger:nicolaides:erlebacher:hussaini,bagheri:brandt:henningson,Semeraro:13}. 
%

Our expression for the control of the baseflow modification leads to a LQ controller that is equivalent to a proportional controller in the case of a single mode, associated with the decay of the baseflow modification. Our approach can be extended to include a larger part of the stable spectrum, for instance including all the modes associated with eigenvalues located along the real axis. In the same perspective, modes with a very small frequency could be included as well, to compute the open loop strategy. In addition, the present method could be used in a closed loop setup, considering only the very slow dynamics, using a Kalman filter and a reduced order model of the baseflow modification. The open-loop strategy of the baseflow could be redesigned by solving the differential Riccati equation where the decay rate of the baseflow modification
$\sigma$ 
could be computed as a function of the initial suction amplitude $\Phi_0$. Methods such as nonlinear control and speed control \citep{geering} where the cost function depends on the amplitude of the baseflow modification are also interesting alternatives.

In the scope of extending the present work to the three-dimensional configuration,
both upstream disturbances and the effect of increasing Reynolds numbers could be addressed by designing a control strategy that modifies the structure of the recirculation region in the spanwise direction $z$. Indeed, \citet{boujo:fanni:gallaire} showed that 
flow modifications in the spanwise direction can decrease transient growth in a shear layer. Such alternatives will probably have to be considered for a more robust control strategy 
when considering a three-dimensional flow geometry.

\Blue{The present strategy could be tested in an experiment, first imposing a suction at the summit of the bump and second varying the decay rate of the suction. At least in the near critical case, the present slowly decaying suction strategy should improve the stability of the system and delay the transition to turbulence beyond the reattachment region.}

\subsection*{Acknowledgments}
This work has been carried out in the framework of the Labex MEC (ANR-10-LABX-0092) and of the A*MIDEX project (ANR-11-IDEX-0001-02) funded by the "Investissements d'Avenir" French Government program managed by the French National Research Agency (ANR). We would also like to acknowledge the anonymous referees for their through reviews and constructive suggestions.

\appendix

\section{Dynamic mode decomposition}\label{DMD_A}

\Blue{In this section we provide the steps used to perform the Dynamic Mode Decomposition in \S\ref{sec:DMD_step}.}
Following the algorithm of \citet{schmid:2010}, snapshots \Blue{of the velocity $\mathbf{U}(\mathbf{x},t)$} are stored for instance at every $\Delta t=3$ such that
\begin{equation}
 \mathbf{L}_1^n = \left[\mathbf{U}(\mathbf{x},t+\Delta t),
 \mathbf{U}(\mathbf{x},t+2\Delta t), \dots,\mathbf{U}(\mathbf{x},t+n\Delta t)\right],
\label{krylov_sequence}
 \end{equation}
where $n=600$ is the number of snapshots stored in the observations matrix $\mathbf{L}_1^n$ and considered in the present study. 
Here the subscript is the number of the first snapshot stored in the observation matrix and 
the superscript is the number of the last snapshot stored in the sequence. To extract the dominant 
features of the sequence (\ref{krylov_sequence}), the $n^{th}$ vector is expressed as a function
of the previous realizations, namely
\begin{equation}
\mathbf{U}(\mathbf{x},t+n\Delta t) = a_1\mathbf{U}(\mathbf{x},t+\Delta t) 
                                 + a_2\mathbf{U}(\mathbf{x},t+2\Delta t)
                         + \dots + a_{n-1}\mathbf{U}(\mathbf{x},t+(n-1)\Delta t) + \mathbf{r},
\end{equation}
\Blue{where $\mathbf{r}$ is the residual vector}, which in matrix form reads
\begin{equation}
\mathbf{U}(\mathbf{x},t+n\Delta t) = \mathbf{L}_1^{n-1}\mathbf{a} + \mathbf{r},
\end{equation}
where $\mathbf{a}^T=[a_1,a_2,\dots,a_{n-1}]$. 
As shown by \citep{Ruhe}, the transition matrix $\mathbf{S}$, defined by
\begin{equation}
 \mathbf{L}_{2}^{n} = \mathbf{L}_{1}^{n-1}\mathbf{S} + \mathbf{r}\mathbf{e}^T_{n-1},
\end{equation}
is of companion type and $\mathbf{e}^T_{n-1}$ is the $(n-1)^{th}$ unit vector.
The only unknowns are the coefficients $\mathbf{a}$, stored in the last column of $\mathbf{S}$.
Rather than computing the $a$ coefficients using for instance, a least squares procedure 
\citep{rowley:mezic}, which can lead to an ill-conditioned companion matrix $\mathbf{S}$ \citep{schmid:2010}, 
it is of advantage to regularize the matrix $\mathbf{S}$ using a Proper Orthogonal Decomposition (POD), 
by performing a Singular Value Decomposition (SVD) of $\mathbf{L}_{1}^{n-1}$ \citep{schmid:2010}
such that
\begin{equation}
\mathbf{L}_{1}^{n-1} = \mathbf{V}\Sigma\mathbf{W}^H,\quad \mbox{ resulting in } \quad
\mathbf{\tilde{S}}= \mathbf{V}^H \mathbf{L}_{2}^{n}\mathbf{W}\Sigma^{-1},
\label{regul}
\end{equation}
where the superscript $^H$ denotes the conjugate transpose. Note that  only the non-zero singular 
vectors $\mathbf{v}$,$\mathbf{w}$, associated with singular values $\varepsilon\ge10^{-8}$ are retained in (\ref{regul}),
for the projection onto the POD modes $\mathbf{V}$.\\
The modal structures of the DMD modes $\varphi$ are extracted from the matrix $\mathbf{\tilde{S}}$
such that
\begin{equation}
 \varphi_k = \mathbf{V}\mathbf{y}_k,
\end{equation}
with $\mathbf{y}_k$, the $k^{th}$ eigenvector of $\mathbf{\tilde{S}}$ such that 
$\mathbf{\tilde{S}}\mathbf{y}_k = \mu_i\mathbf{y}_k$ and $\mathbf{V}$ are
the left singular vectors computed in (\ref{regul}). The continuous time eigenvalues $\sigma_k$ are 
computed using the transformation
\begin{equation}
\sigma_k=\log(\mu_k)/\Delta t,
\end{equation}
where the $\log$ is the complex logarithm.

\section{Linear quadratic controller}\label{sec:appendix_LQR}

\Blue{In this section, we provide the main steps to derive the Riccati equation (\ref{Ricc}). Following the model described in \S\ref{sec:control_bf} for the baseflow modification dynamics subject to a time dependent suction, one gets the first order system}
\begin{subeqnarray}
\frac{\mbox{d}\, \delta A_{\mbox{rec}}(t)}{\mbox{d}t} &=& \sigma \delta A_{\mbox{rec}}(t) + \mathcal{G}\Phi(t), \\ 
\Phi(t)&=&K\delta A_{\mbox{rec}}(t).
\label{mean-lin-sys_app}
\end{subeqnarray}
The aim is to find a control law $\Phi(t) : [0, \infty)$, such that the controllable system (\ref{mean-lin-sys_app})(a-b) is transferred from an initial state $\delta A_{\mbox{rec}}(t=0)$, to an arbitrary final state 
$\delta A_{\mbox{rec}}(t\rightarrow\infty) = 0$ such that the cost functional (\ref{JOL})
is maximized and attains a finite optimal value. The time evolution of the control system (\ref{mean-lin-sys_app}(a-b) and the cost function (\ref{JOL}) are ruled by the Hamiltonian
\begin{equation}
\mathcal{H} = \delta A_{\mbox{rec}}^2(t) + \gamma\Phi^2(t) 
+ Z(\mathcal{G}\Phi(t) + \sigma \delta A_{\mbox{rec}}(t)),
\label{hamiltonian}
\end{equation}
where a co-state variable $Z(t)$ has been introduced.
Taking derivatives prior to the state, the co-state and the control yields the necessary conditions for optimality \citep[see][pg. 41-42]{geering} such that
\begin{subeqnarray}
\nabla_{Z} \mathcal{H} &=& \frac{\mbox{d}\delta A_{\mbox{rec}}}{\mbox{d}t} = \sigma \delta A_{\mbox{rec}} + \mathcal{G}\Phi, \\
\nabla_{\delta A_{\mbox{rec}}} \mathcal{H} &=& \frac{\mbox{d}Z}{\mbox{d}t} = \sigma Z + \delta A_{\mbox{rec}},\\
\nabla_{\Phi} \mathcal{H} &=& \gamma\Phi + Z\mathcal{G},\\
\delta A_{\mbox{rec}}(t_0) &=& \mathcal{G}\Phi_0.
\label{khun-tucker-cond}
\end{subeqnarray}
Minimizing the Hamiltonian (\ref{hamiltonian}) corresponds to canceling each derivative in (\ref{khun-tucker-cond})(a-c). Canceling (\ref{khun-tucker-cond})(c) yields the optimal open-loop control law
\begin{equation}
\Phi=-\gamma^{-1} \mathcal{G} Z. 
\end{equation}
Substituting this control in the evolution equations (\ref{khun-tucker-cond})(c) results in a two-point initial value problem
\begin{subeqnarray}
\frac{\mbox{d}\delta A_{\mbox{rec}}}{\mbox{d}t} &=& \sigma \delta A_{\mbox{rec}} + \mathcal{G}\Phi, \\
\frac{\mbox{d}Z}{\mbox{d}t} &=& \sigma Z + \delta A_{\mbox{rec}},\\
\Phi&=&-\gamma^{-1} \mathcal{G} Z,\\
\delta A_{\mbox{rec}}(t_0) &=& \mathcal{G}\Phi_0,\\
Z(t_1) &=& \delta A_{\mbox{rec}}(t_1).
\label{dir-ad}
\end{subeqnarray}
Introducing the linear ansatz
\begin{equation}
Z(t) = X(t)\delta A_{\mbox{rec}}(t),
\label{ansatz2}
\end{equation}
differentiating (\ref{ansatz2}) with respect to time, and making use of equations (\ref{dir-ad})(a-c) reduces to a single equation
\begin{equation}
\begin{split}
\delta A_{\mbox{rec}}\frac{\mbox{d}X}{\mbox{d}t} +  X\frac{\mbox{d}\delta A_{\mbox{rec}}}{\mbox{d}t} =  \delta A_{\mbox{rec}}\frac{\mbox{d}X}{\mbox{d}t} + X(\sigma-\gamma^{-1}\mathcal{G}^2X)\delta A_{\mbox{rec}} = -\delta A_{\mbox{rec}} - \sigma X\delta A_{\mbox{rec}},
\end{split}
\label{ansatz2time}
\end{equation}
which is the differential Riccati equation
\begin{equation}
\left(-\frac{\mbox{d}X}{\mbox{d}t} -\gamma^{-1} X^2 \mathcal{G}^2 + 2\sigma X + 1\right)\delta A_{\mbox{rec}}\equiv 0.
\label{riccati}
\end{equation}
Because equation (\ref{mean-lin-sys}) has constant coefficients,
the time dependence can be dropped and equation (\ref{riccati}) reduces to the algebraic Riccati equation (\ref{Ricc}).

\bibliographystyle{jfm}

\bibliography{bib}

\end{document}